%% file: mnras_template.tex
\documentclass[fleqn,usenatbib]{mnras}
\usepackage{natbib}
\usepackage{xspace}
\usepackage{amsmath}

\newcommand{\logg}{\ensuremath{\log g \,\,}}

\newcommand{\feh}{\ensuremath{\protect\rm [Fe/H] } }
\newcommand{\cfe}{\ensuremath{\protect\rm [C/Fe] }}
\newcommand{\teff}{\ensuremath{T_{\mathrm{eff}}}\xspace}
\newcommand{\msun}{\rm M_{\odot}}
\newcommand{\Zsun}{\rm Z_{\odot}}
\usepackage{graphicx}
\usepackage{tablefootnote}

\usepackage{footnote}
\usepackage{amsmath}
 \usepackage{subcaption}
\usepackage{color}
\usepackage{xcolor}

\usepackage{hyperref}

\usepackage{deluxetable}
\usepackage{longtable}
\usepackage{newtxtext}

\usepackage[T1]{fontenc}
\usepackage{ae,aecompl}

\usepackage{graphicx}	
\usepackage{amsmath}	
\usepackage{amssymb}	

\title[]{PISN-explorer: hunting the descendants of very massive first stars\thanks{Based on observations made with the ESO Very Large Telescope at the La Silla Paranal Observatory under program ID 108.23N5.001}}

\author[D. Aguado et al.]{D.~S. Aguado$^{1,2,5}$\thanks{E-mail: david.aguado@unifi.it},
S. Salvadori$^{1,2}$,
\'A. Sk\'ulad\'ottir$ ^{1,2}$, E. Caffau$^{3}$, P. Bonifacio$^{3}$,
\newauthor I. Vanni$^{1,2}$, V. Gelli$^{1,2}$, I. Koutsouridou$^{1,2}$, and A. M. Amarsi$^{4}$  
\\
\\
$^{1}$Dipartimento di Fisica e Astronomia, Universit \'a degli Studi di Firenze, Via G. Sansone 1, I-50019 Sesto Fiorentino, Italy \\
$^{2}$INAF/Osservatorio Astrofisico di Arcetri, Largo E. Fermi 5, I-50125 Firenze, Italy\\
$^{3}$GEPI, Observatoire de Paris, Université PSL, CNRS, 5 Place Jules Janssen, 92190 Meudon, France.\\
$^{4}$Theoretical Astrophysics, Department of Physics and Astronomy, Uppsala University, SE-751 20 Uppsala, Sweden\\
$^{5}$Instituto de Astrof\'{\i}sica de Canarias,
              V\'{\i}a L\'actea, 38205 La Laguna, Tenerife, Spain\\
}

\date{Accepted XXX. Received YYY; in original form ZZZ}

\pubyear{2022}

\begin{document}

\maketitle

\begin{abstract}
 The very massive first stars ($m>100$\,$\msun$) were fundamental to the early phases of reionization, metal enrichment, and super-massive black hole formation. Among them, those with $140\leq\rm m/\msun\leq260$ are predicted to evolve as Pair Instability Supernovae (PISN) leaving a unique chemical signature in their chemical yields. Still, despite long searches, the stellar descendants of PISN remain elusive. Here we propose a new methodology, the {\tt PISN-explorer}, to identify candidates for stars with a dominant PISN enrichment. The {\tt PISN-explorer} is based on a combination of physically driven models, and the {\tt FERRE} code; and applied to data from large spectroscopic surveys (APOGEE, GALAH, GES, MINCE, and the JINA database). We looked into more than 1.4 million objects and built a catalogue with 166 candidates of PISN descendants. One of which, 2M13593064+3241036, was observed with UVES at VLT and full chemical signature was derived, including the \textit{killing} elements, Cu and Zn. 
 We find that our proposed methodology is efficient in selecting PISN candidates from both the Milky Way and dwarf satellite galaxies such as Sextans or Draco.
 Further high-resolution observations are highly required to confirm our best selected candidates, therefore allowing us to probe the existence and properties of the very massive First Stars.

\end{abstract}




\section{Introduction}
The first (Pop\,III) stars formed out of primordial composition gas, i.e. in environments where the fragmentation process was less efficient then in local star-forming regions, thus enabling the formation of more massive stars than those that we observe today: from a few tens, up to {\it thousands} of solar masses \citep[e.g.][]{susa14,hirano2015}. Their characteristic mass, furthermore, was likely $\geq 10$\,$\msun$ \citep[e.g.][]{Bromm13}. These theoretical findings have strong physical grounds and have been reported across decades by means of different analytical calculations \citep[see e.g.][]{silk1977, mckee2008}, numerical simulations \citep[see e.g.][]{abel2002, hosokawa2011}, and studies of stellar archaeology \citep[e.g.][]{ishigaki18,Rossi+21}. According to cosmological models, these primitive stars were likely hosted by the bulge and the stellar halo of the Milky Way  \citep[see e.g.][]{tum10, salvadori10, sta17II} and by Local group dwarf galaxies \citep[e.g.][]{salv15}.  From an observational point of view, however, we are still lacking key probes of first stars in the high-mass regime ($\rm m_*>140$\,$\rm M_\odot$) very recently some indirect hints have been provided (Pagnini et al. sumbitted). 


Very massive first stars, $140 \leq \rm m_*/M_{\odot} \leq 260$, are predicted to end their lives as energetic Pair Instability Supernovae (PISN), which inject  $50\%$ of their mass in the form of metals into the interstellar medium (ISM). Thus, a single PISN will strongly enrich the primordial gas with a probability distribution function peaking at $\mathrm{[Fe/H] \approx-2.0}$ \citep[][]{karlsson2008, bennassuti2017, salvadori19}, leaving a unique chemical signature with a strong odd-even effect \citep[e.g.][]{heger2002, tak18}.
Unfortunately, pure PISN descendants are predicted to be extremely rare: even at the peak metallicities, $\mathrm{[Fe/H] \approx-2.0}$, they are thought to represent $<0.1\%$ of Milky Way stars \citep{bennassuti2017}. The traditional approach of searching only for stars showing 100\% PISN enrichment has thus not proven successful. Rather, a large fraction of PISN descendants are expected to also have a significant contribution from normal Pop\,II stars, which can form early on in the Pop\,III-enriched ISM and then evolve as core-collapse supernovae (CCSN). 

The abundances of the so-called \textit{killing} elements Cu and Zn \citep{salvadori19} may be the smoking gun for a true PISN descendant. These elements are barely produced by PISNe,  but yielded by all other supernovae (see also Vanni et al. in prep). The extreme sub-solar abundances of Zn and Cu with respect to Fe from PISNe, persist even in environments polluted by other sources up to a $50\%$ level \citep{salvadori19}. But given the rarity of such stars, until now, only two descendants of PISN have been reported \citep{aoki14, salvadori19}. We are thus in the situation where even a few more identified PISN descendants would greatly advance the field.

In the recent years there has been a breakthrough in large spectroscopic surveys targeting stars in and around the Milky Way (e.g.~Gaia, APOGEE, GALAH, GES). Millions of stars have been observed with intermediate- to high-resolution spectra ($R\gtrsim10,000$), and these numbers will further increase with two large upcoming spectroscopic surveys, starting operations in the next two years: WEAVE in the Northern hemisphere \citep{Dalton2016}, and 4MOST in the South \citep{deJong2019}. For the first time, we are thus able to have the statistics to identify and characterise rare populations, such as PISN descendants. 

Searching through such large databases, however, requires a focused and dedicated effort, and with this goal in mind we have developed an innovative methodology, the {\tt PISN-explorer}. With a combination of theoretical models \citep{salvadori19}, and the {\tt FERRE} code \citep{alle06}, the {\tt PISN-explorer} exploits all the chemical abundances measured by large surveys (e.g. APOGEE, GALAH, GES, or MINCE) to identify stars that have likely been dominantly enriched by PISN ($\gtrsim50\%$). Here we present this new method, allowing us for the first time to use large databases of chemical abundances to systematically search for the elusive PISN descendants.




\section{The theoretical models}\label{sec:models}
The models used by the {\tt PISN-explorer} are adopted from the general parametric study presented in \cite{salvadori19}. 
This study provides predictions for the chemical properties of an ISM predominantly imprinted by very massive first stars exploding as PISNe, i.e. where the PISNe products account for $\geq 50\%$ of metals in the ISM. The model is general because it condenses the unknown physical processes related to early cosmic star-formation, metal diffusion, and mixing, into three free parameters: 1)~the star formation efficiency, $f_*$,  which provides the fraction of ISM gas condensed into stars; 2)~the dilution factor, $f_{\rm dil}$,   which parametrises the amounts of metals effectively retained into the ISM; 3)~and the mass fraction of ISM metals contributed by PISNe, $f_{\rm PISN}$. Predictions for a PISN-imprinted ISM are provided by exploring the full parameter space. Hence the predictions are general and essentially model independent. 

The ansatz of this approach is to assume that a single very massive first star exploding as PISN can form in the primordial star-forming (mini-)haloes. This assumption is strongly supported by the results of hydrodynamical cosmological simulations following the formation of the first stars \citep[e.g.][]{Hirano2014}. The chemical enrichment of the ISM is evaluated after the injection of metals by a single PISN ($f_{\rm PISN}=100\%$) with different progenitor masses, $m_{\rm PISN}=[140-260]\mathrm M_{\odot}$. It turns out that the final metallicity (or [Fe/H]) of the ISM is settled by the PISN mass and by the ratio between the star-formation efficiency and the dilution factor, $f_*/f_{\rm dil}$. Furthermore, it is always $Z_{\rm ISM} > 10^{-3}\Zsun$, which implies that ``normal'' Pop\,II stars can form out of this medium and thus contribute to the subsequent ISM enrichment (see Fig.~2 of \citealt{salvadori19}). 

The abundance ratios of each element X (from C to Zn) with respect to iron, are computed by varying the relative contribution of PISN and Pop\,II stars to the chemical enrichment ($f_{\rm PISN}=[50-100]\%$) and by assuming that Pop\,II stars form according to a standard  Larson Initial Mass Function (IMF):  $\phi(m_{\star}) = m^{-2.35}_{\star} exp(- m_{ch}/m_{\star})$ with $m_{ch}=0.35 \msun$ \citep{Lars98}. The calculations are made by adopting the yields by \cite{heger2002} for very massive first stars exploding as PISNe, ${\rm {Y^{\rm PISN}_{\rm X}}}(m_{\rm PISN})$, and of \cite{Woosley1995} for Pop\,II stars with initial masses $m_*=[8-40]\mathrm M_{\odot}$ and metallicities, $Z_*=[10^{-4};1]\mathrm Z_{\odot}$, which explode as CCSN, ${\rm Y_X}^{\rm II}(m_*,Z_*)$. 
According to \cite{Woosley1995} we assume that stars between 40 and 140 $\msun$ collapse directly into a black hole thus not
contributing to the chemical enrichment. Note that since we are integrating over the entire Pop\,II IMF the derived yields are not so different from those obtained with models assuming that $m_* > 20\msun$ stars produce a negligible amount of heavy elements \citep[e.g.][]{Limongi+18}. 
\cite{salvadori19} demonstrated that the ISM abundance ratios, [X/Fe], or the chemical abundance pattern of the stars formed out of it, depend upon $f_{\rm PISN}$, the yields of PISN and of Pop\,II stars; but they are not directly affected by $f_*/f_{\rm dil}$ although this ratio controls the metallicity of the Pop\,II stars contributing to the ISM enrichment.

\begin{figure*}
  \centering
  \caption{Example of the {\tt PISN-explorer} analysis performed on an APOGEE star. }
     \begin{subfigure}[b]{0.48\linewidth}
         \centering
         \includegraphics[width=0.73\hsize, angle=90]{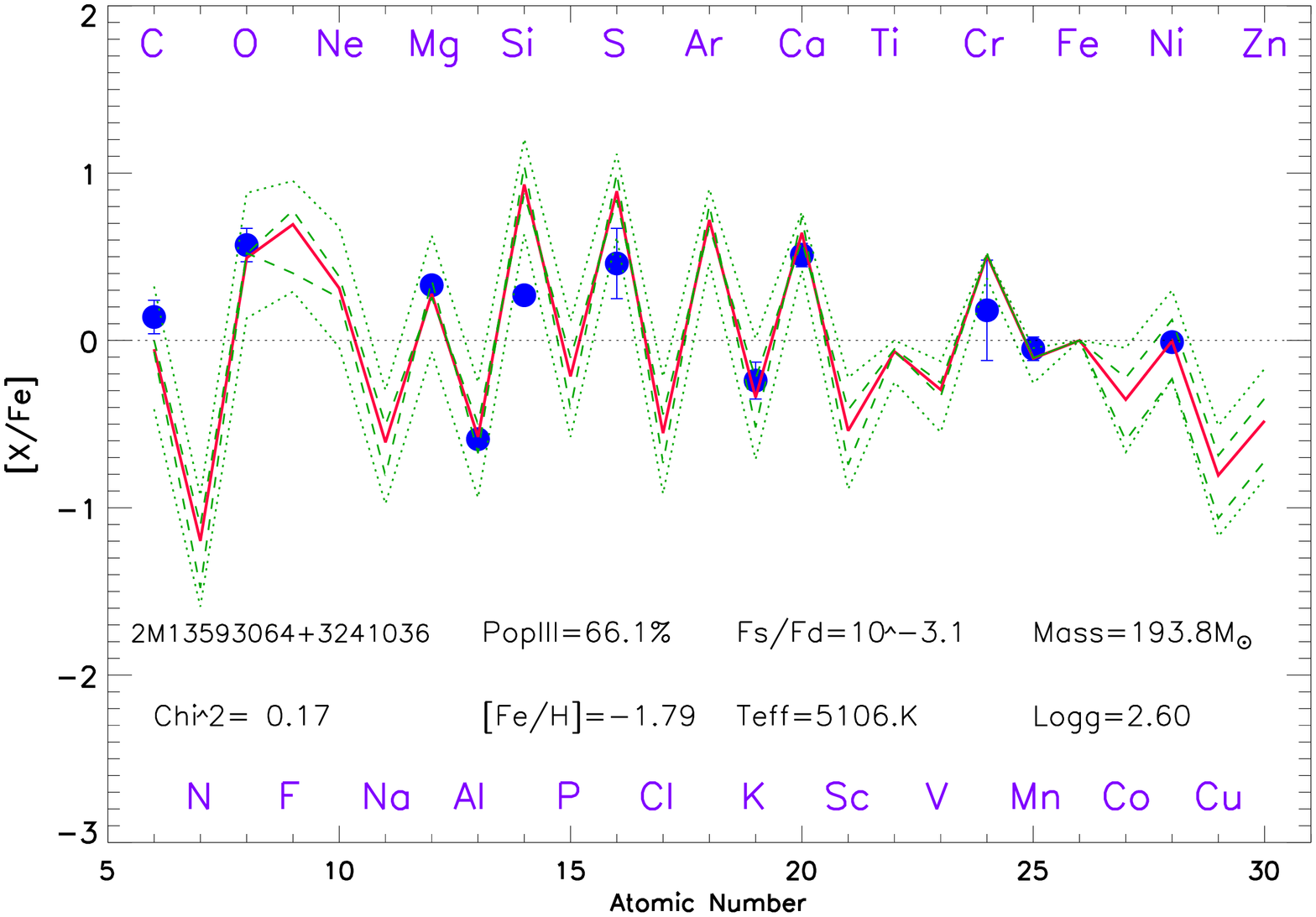}
         \caption{ The measured chemical signature of 2M13593064+3241036 from APOGEE (blue dots) together with the best fit to a PISN model computed with the {\tt PISN-explorer} (red line).  Derived PISN parameters and main stellar parameters are also shown. Finally, for comparison we show interpolated models with $\pm 20\msun$ of PISN mass (dotted lines) and $\pm 15\%$ of PISN contamination (dashed lines).}
         \label{fig:example}
     \end{subfigure}
     \hfill
     \begin{subfigure}[b]{0.48\linewidth}
         \centering
         \includegraphics[width=1.05\hsize, angle=0]{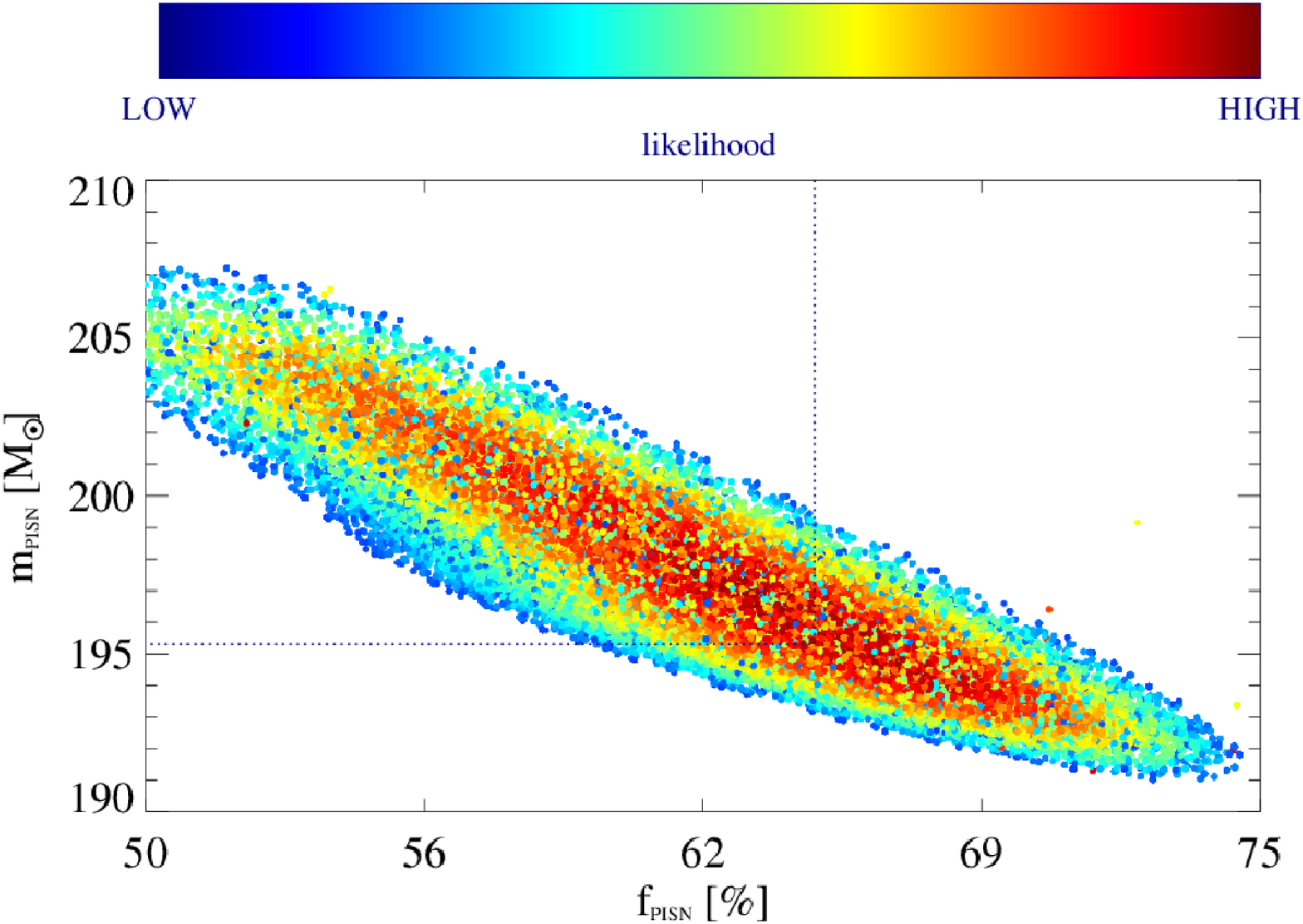}
         \caption{ $f_{\rm PISN}-m_{\rm PISN}$ space of solutions when applying and MCMC algorithm computed with {\tt FERRE} over the chemical signature of 2M13593064+3241036 including 10 chains of 50,000 experiments each. Color bar represents the likelihood of each solution.  
         \newline
         \newline}
         \label{fig:mcmc}
     \end{subfigure}
\end{figure*}

\section{Exploited datasets}\label{sec:datasets}
The goal of this work is to efficiently select candidates for PISN descendants from existing and publicly available data. Suitable databases should provide reliable elemental abundances for FGK stars together with stellar parameters: effective temperature, surface gravity, and metallicity (\teff, \ensuremath{\log g}, and [Fe/H]). From the available spectroscopic surveys, we restrict our selection to the large surveys that have the best chemical abundances information (APOGEE, GALAH, Gaia-ESO, and the MINCE survey). In addition, we exploit high-quality literature data available through the JINA database. Here, we will mainly focus on elements from C to Zn but sometimes abundances of heavier elements (Ba, Ce) could be also used to impact the selection, since these are expected to be low in PISN descendants. Finally, in the following we assumed the solar abundances each survey used at that time. Future -improved- measurements of solar abundances are easily applicable to our methodology.

\subsection{APOGEE}
The Apache Point Observatory Galactic Evolution Experiment~2 \citep[APOGEE-2;][]{apogee17} is a near-infrared high-resolution survey in the $H$-band (1.514-1.696\,$\rm \mu m$ at $R\sim22,500$) that provides stellar parameters (\teff, \logg, [Fe/H]), radial velocities, and elemental abundances. Solar abundances are those from \citet{asp05}. The high quality of APOGEE measurements at metallicities down to $\rm [Fe/H]\sim-2$ clearly recommends the use of this spectroscopic survey. 
For our purpose, we use the \textit{final allStar} catalogue of APOGEE Data Release (DR) 17~\citep{Abdurrouf_APOGEE2022} which includes up to 20 elemental abundances (C, N, O, Na, Mg, Al, Si, P, S, K, Ca, Sc, Ti, Cr, Mn, Fe, Co, Ni, Cu, and Ce) for 879,437 spectra (733,901 objects)\footnote{The APOGEE catalogue can be retrived here: \url{https://www.sdss.org/dr17/irspec/spectro_data/}}. Following the recommendation of the APOGEE team, we discard both \ion{Ti}{i} and \ion{Ti}{ii} from further analysis. Unfortunately, the Cu absorption lines in the infrared are extremely week and therefore only detectable for metal-rich stars, $\rm [Fe/H]>-1$.

\subsection{GALAH}
The GALactic Archaeology with HERMES \citep[GALAH,][]{galah15} survey is a large spectroscopic survey in the optical at a resolving power $R\sim28,000$, covering four wavelength windows within 4713 to 7887\,\AA. The third data release \citep[DR3;][]{Buder_GALAH2021}, including the \textit{main\_allspec\_v2} catalogue\footnote{GALAH catalogue: \url{https://cloud.datacentral.org.au/teamdata/GALAH/}}, provides for 588,571 stars: stellar parameters, radial velocities, and chemical abundances for up to 30 elements (Li, O, C, Na, Mg, Al, Si, K, Ca, Sc, Ti, V, Cr, Mn, Co, Ni, Cu, Zn, Rb, Sr, Y, Zr, Mo, Ru, Ba, La, Ce, Nd, Sm, Eu). GALAH provides chemical abundances at $\rm[Fe/H]\gtrsim-2$, but not all elements are available at the lowest metallicities. Solar abundances are from \citet{grevesse2007}.

\subsection{Gaia-ESO}
The Gaia-ESO Survey \citep[GES,][]{gaia-eso, gilmore2022} has observed 115,000 stars from the main components of the Milky Way, including star clusters, using FLAMES at the VLT. The employed instruments (GIRAFFE and UVES) provide high-resolution spectra, $16,200<\rm R<47,000$, covering various wavelength ranges, depending on the configuration.
We retrieved\footnote{Gaia-ESO catalogue: \url{https://www.eso.org/qi/catalogQuery/index/393}} the latest available DR5 \citep{gesdr5} with 114,324 stars. The solar abundances used for this survey are those from \citet{asp09}. A significant fraction of GES pointings is dominated by metal-rich stars  $\rm [Fe/H]>-1$, which are less likely to be PISN descendants. These fields can, however, also contain some metal-poor stars, so we included the entire dataset. Including the entire GES sample also ensures that our search is as unbiased as possible. The GES DR5 dataset includes stellar parameters, radial velocities, and chemical abundances for up to 31 elements (He, Li, C, N, O, Na, Mg, Al, Si, S, Ca, Sc, Ti, V, Cr, Mn, Co, Ni, Cu, Zn, Sr, Y, Zr, Mo, Ba, La, Ce, Pr, Nd, Sm, Eu) 

\subsection{MINCE}
 Measuring   at   Intermediate   metallicity Neutron-Capture Elements \citep[MINCE,][]{mince} is a project aiming to study abundances for neutron-capture elements using different facilities such as HARPS, UVES, or FIES. Conveniently, the targets selection is designed to observe intermediate and very metal-poor stars ($\rm -0.7 \gtrsim [Fe/H] \gtrsim -2.5$). For this work we employed the first year sample including 43 stellar targets and high quality elemental abundances O, Na, Mg, Al, Si, S, Ca, Sc, Ti, V, Cr, Mn, Co, Ni, Cu, Zn. The employed solar abundances are those from \citet{lod09} and \citet{caff11II} for O and S. The following data releases in subsequent years will be a valuable source of PISN descendant candidates. 

\subsection{JINA}
The Joint Institute for Nuclear Astrophysics (JINA) database \citep{jina18} is a collection of literature papers with high-resolution analysis of 1658 metal-poor stars  $\rm [Fe/H]<-2$, including a variable range of elemental abundances (from Li to U). We downloaded the whole database\footnote{JINA database: \url{https://jinabase.pythonanywhere.com/}}, set the solar abundances to \citet{asp09}, and treat each entry in the same way regardless of its origin (e.g. halo, bulge, or dwarf galaxy members). 

\begin{figure}
\begin{center}
{\includegraphics[width=60 mm, angle=90,trim={ 0.cm 0.cm 0.cm 0.cm},clip]{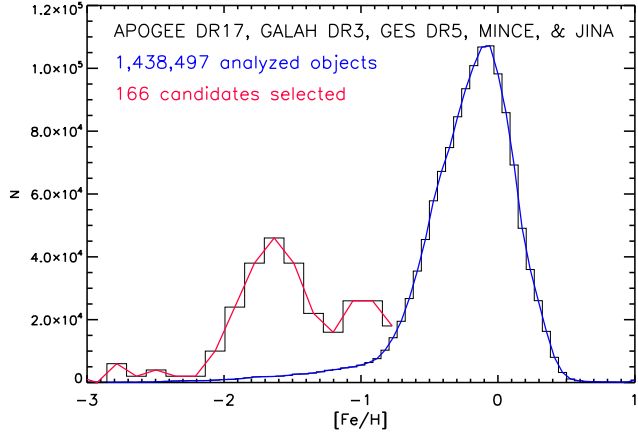}}
\end{center}
\caption{Metallicity distribution functions for the 1.4 million stars analysed (blue), and the 166 PISN selected candidates (red). Note that the red distribution has been arbitrarily scaled for easier comparison.}
\label{fig:mdf}
\end{figure}

\section{Methodology}\label{sec:methodology} 
We have developed the {\tt PISN-explorer}, a systematic methodology to efficiently select PISN-polluted candidates from observational data (Sec.~\ref{sec:datasets}) using the {\tt FERRE} code\footnote{{\tt FERRE} is available from \url{http://github.com/callendeprieto/ferre}} from \citet{alle06} and theoretical predictions from \citet{salvadori19} (Sec.~\ref{sec:models}).
 {\tt FERRE} is a {FORTRAN} code optimised for spectral analysis \citep[e.g.][]{alle14, agu17I} but capable of performing fitting of any class of data to a rectangular grid of theoretical models. The {\tt PISN-explorer} methodology implies several steps:
\begin{enumerate}
    \item \textbf{Packaging of models:}
    A grid of theoretical predictions for the chemical abundances of PISN descendants (50-100\% PISN pollution) is created with a {\tt FERRE}-friendly header. Key information included in the header: a)~the number and label for each free parameter included in the models ($f_{\rm PISN}$, ${\log_{10}{(f_*/f_{\rm dil}})}$, and $m_{\rm PISN}$); b)~the minimum value for each parameter (0.5, -4, and 140\,$\msun$ respectively); c)~the steps for each parameter (0.1, 1, 9$\msun$, respectively); d)~the number of elements (``pixels'') within the models (25, counting from C to Zn), and their atomic numbers (``$\lambda$-array''). Finally, the grid is complete with the predictions ($\rm [X_{i=6,30}/Fe]$) in rows just below the header.  Here, as explained in Sec. \ref{sec:models}, we use Pop\,II yields from \citep{Woosley1995} to be fully consistent with \citet{salvadori19}. However, if the potential user of {\tt PISN-explorer} wishes to use a different set of models \citep[e.g.][or any other that can be published in the future]{Nomoto2013,Limongi+18} the grid has to be re-computed as explained along these lines. 
    \\
    \item \textbf{Preparing the data:} Following the recommendation of each survey, we discard stars with suboptimal flags, {\tt\string STAR\_BAD} {\tt\string ASPCAPFLAG} for APOGEE and {\tt\string flag\_sp}$\neq0$ for GALAH, while all the sample was taken from Gaia-ESO, MINCE, and JINA. Additionally, we only considered elemental abundances with {\tt\string x\_fe\_flag} $= 0$ in both APOGEE and GALAH, meaning that only reliable values are considered. 
    For each star in our sample we prepared three different files with 25-columns length each: 1)~the measured chemical ratios ($\rm [X_{i=6,30}/Fe]_{obs}$) from C to Zn; 2)~the uncertainties ($\sigma_{\rm [X_{i=6,30}/Fe]_{obs}}$); and 3)~the weight we gave to each measurement ($\omega_{\rm [X_{i=6,30}/Fe]_{obs}}$). The $\omega$ values are equal to 0 if there is no measurement for element X, 1 for Cu and Zn, and 0.5 for the rest. Thus, we give double weight to the \textit{killing} elements since they are the smoking gun of PISN pollution. This tuning implies that the  $\chi^{2}_{0}$ will heavily depend on the Cu and Zn if they are available but no effect otherwise.
        \\

    \item \textbf{Launching the code:} {\tt FERRE} is able to look for the best fit by interpolating between the nodes of the grid. We used quadratic interpolation and Nash's truncated Newton algorithm \citep{NASH90}. After the minimum $\chi^{2}{0}$ is found, the code produces the set of best parameters with errors, along with an interpolated model. In Fig. \ref{fig:example} we show an example of a fit with the observed values $(\rm [X_{i=6,30}/Fe]_{obs}$, blue dots) from the literature for a single APOGEE object, and the best fit $ (\rm [X_{i=6,30}/Fe]_{fit}$, red line) derived with {\tt FERRE}, giving a set of parameters: $f_{\rm PISN}=66.1$\%; $f_*/f_{\rm dil}=10^{-3.1}$; and $m_{\rm PISN}=193.8\msun$. We also calculated a reduced $\chi^{2}$ as $\chi^{2}=\chi^{2}_{\circ}/(N-M)$, where $\chi_{\circ}^{2}$ is the one defined in Equation 14 of \citet{salvadori19}, N the number of fitting points, and M the number of free parameters of the model. Therefore, we will have reduced $\chi^{2}$ values $>0$. In the following we will refer to this reduced $\chi^{2}$.
            \\

    \item \textbf{Interpreting the results and caveats:} The results of the {\tt PISN-explorer} are interpreted, i.e. the best set of parameters ($f_{\rm PISN}$, ${\log_{10}{(f_*/f_{\rm dil}})}$, and $m_{\rm PISN}$), and an interpolated model ($\rm [X_{i=6,30}/Fe]_{fit}$). When a star's chemical signature is very far from the theoretical predictions (the vast majority of the cases), the code will not be able to find a good fit and will usually end up on a solution with a high $\chi^2$ at the limit of the grid ($m_{\rm PISN}=140$\,$\msun$, for example). This is because the code is trying to find something beyond the ranges of the grid, and should therefore be interpreted as a non-reliable solution. Unless the quality of the fit is very high, solutions with $f_{\rm PISN}$ and $m_{\rm PISN}$ close to the limit of the grid are discouraged. However, the ${\log_{10}{(f_*/f_{\rm dil}})}$ parameter is less sensitive to model-to-model variations because the predicted abundance patterns do not directly depend upon this quantity (see Sec.~\ref{sec:models}). Therefore good fits can appear at the limits of the ${\log_{10}{(f_*/f_{\rm dil}})}$ range. Objects with a very low number of derived elemental abundances ($\leq5$) lead to no recommended solutions. Additionally, there are some elements that are more affected by stellar processes in evolved stars. C, for instance, is  converted into N during the CN cycle, and brought to the surface during the red giant branch phase. Thus, it is highly recommended to correct for these stellar evolutionary effects \citep[see, e.g.][]{pla14}. Finally, any available corrections due to inhomogeneities in the three-dimensional (3D) stellar atmosphere and the non-local thermodynamic equilibrium (NLTE) state of the matter for elemental abundances will significantly improve the results of the analysis.
        \\

    \item \textbf{Looking for asymmetric signatures:}   The well known odd-even bias is a remarkable feature of measured stellar chemical compositions \citep[see, e.g.][]{payne25,russel41}. Elements with an even atomic number (C, O, Mg, Si, Ca, Ti) are more easily detected in the atmospheres of metal-poor stars and therefore most of the surveys are able to provide them for a large number of objects with good precision. Consequently, the available chemical signatures are asymmetric, i.e. biased towards even-elements. This is unfortunate, since a stark contrast between the abundances of odd and even elements is a key signature of PISN yields (see e.g.~\ref{fig:example}). Therefore, the higher is the number of available odd elemental abundances the better performance can be obtained with this methodology.
        \\

    \item \textbf{MCMC validation:} To ensure that degeneracy in the parameters determination is not affecting our methodology we used a Markov-Chain Monte-Carlo (MCMC) analysis based on self-adaptive randomised subspace sampling \citep{vru09}. This algorithm is also implemented in the {\tt FERRE} code and can help to understand the solution space, and offer a statistical validation for the evaluation of uncertainties. We launched 10 chains of 50,000 experiments each and let the code burn up to 500 experiments in each chain. In Fig. \ref{fig:mcmc} we show the likelihood in the $f_{\rm PISN}-m_{\rm PISN}$ space of solutions. Although the employed algorithm looking for the best solution is different, unsurprisingly the best solution is compatible within the uncertainties with the one shown in Fig. \ref{fig:example} ($f_{\rm PISN}=65$\%; $f_*/f_{\rm dil}=10^{-3.1}$; and $m_{\rm PISN}=195.3$\,$\msun$). The likelihood distribution (Fig.~\ref{fig:mcmc}) shows solution areas with similar probability that are smaller than the step of the grid.  Furthermore, the smooth behaviour of the likelihood distribution shows that no evident degeneracies are playing a critical role in our calculations. Therefore, this MCMC test demonstrates that the solution {\tt FERRE} is finding is stable and the nodes corresponding to the space parameters are small enough to account for the required sensitivity.
    \end{enumerate}

\section{Target Selection}\label{sec:selection}
Our sample mainly consists of nearby stars from the disk, the bulge, and the halo of the Milky Way (Sec.~\ref{sec:datasets}). We analyzed 1,438,497 stars following the methodology explained in Sec. \ref{sec:methodology} with no further cuts applied at this stage. In Fig. \ref{fig:mdf} we show in blue the metallicity distribution function of the whole sample, including GALAH, APOGEE, GES, and JINA stars with available measurements of chemical abundances.
The target selection based on the {\tt PISN-explorer} analysis is a two step procedure based first on hard cuts applied blindly over the entire sample, {\it systematic selection}, and secondly on more specific and survey-dependent selection, {\it individual selection}. We finally end up with our golden catalogue of candidates for further high-resolution follow-up. The three stages are detailed below:

\begin{itemize}
    \item \underline{Systematic Selection.}  The MDF for the entire considered sample is shown in red in Fig. \ref{fig:mdf}.
    Following the known caveats (discussed in Sec.~\ref{sec:methodology}), we discarded objects with less than 5 available chemical abundances. In addition, since we focus on FGK type metal-poor stars we only include stars with $4000$\,K$<$\teff$<7000$\,K and \feh$<-0.7$. Hotter stars have much weaker metallic absorption lines, and it is in general not possible to derive a complete chemical signature. On the other hand, the early M type stars are dominated by molecular bands, and thus cooler stars would also prevent from determining accurate chemical abundances. 
    The remaining sample of 385202 stars is run through the {\tt PISN-explorer}.  Subsequently, we discard fits with $\chi^2>0.35$ and remove out after inspection solutions with $m_{\rm PISN}<145\msun$ and/or $f_{\rm PISN}<55$\% since they are close to the limit of the grid.  
    
    \item \underline{Individual Selection.}   Hard cuts themselves do not guaranty an optimal selection.  First of all, the \textit{killing} elements, Cu and Zn are important guidelines and, if available, we discard objects with $\rm [Cu,Zn/Fe]>0$. In addition, we maximised, when possible, odd-even effect by selecting objects with $\rm [Mg/Al]>0.7$. We also preferred objects with $\rm [Mg/Ca]<0.0$ that is also a desirable sign according to our models. After applying all of this criteria we end-up in a catalogue of 166 candidates from all included surveys (Sec.~\ref{sec:datasets}).  In Fig. \ref{fig:mdf} we also show in red our selection. It is quite clear that different datasets contribute in a different manner. While the most metal-poor bump ($\feh\lesssim-2.5$) is mainly from the JINA database, the most metal-rich peak ($\feh\approx-1$) is dominated by GES stars. However, the bulk of our selection based on GALAH and APOGEE is peaked at $\rm [Fe/H]=-1.7$, close to the most likely metallicity region where PISN descendants are predicted to live \citep[e.g.][]{karlsson2008,salvadori19}. We note that the red MDF showed in Fig. \ref{fig:mdf} is, at first order, independent of the selection criteria we applied.
    
    \item \underline{Golden catalogue}  We included extra criteria before consider any star for future follow-up. At this stage more careful direct inspection is recommended, analysing the suitability of each candidate individually. Therefore, we included two objects that did not completely achieve the general criteria $\chi^2<0.35$ but have $\rm [Zn/Fe]<-0.8$. In addition to it, we removed objects with clear s-process enrichment (Ba or Ce) to discard possible contribution from AGB stars. In addition to it, to select objects with large number of abundances (N$ \geq 10$) is obviously a desirable strategy so we prioritise those to others with better fit but less complete chemical information. Finally, we also include suitable observability cuts (V$_{mag}\leq$ 16) ending up in a 45 objects golden catalogue that contains the most promising candidates to be enriched by PISN pollution.  
\end{itemize}

Following the selection criteria described in this section we ended up with our golden catalogue of 45 objects\footnote{The catalogue will be published in a forthcoming paper including a set of new high-resolution observations.}. One of the most promising candidates, 2M13593064+3241036 from APOGEE, is shown in Fig. \ref{fig:example}. For this object the quality of the fit is remarkable: measurements for two odd elements were available (Al and K), and the odd-even effect is clear, $\rm [Mg/Al]=+0.92$,  $\rm [Mg/Ca]<-0.18$, and  $\rm [Ca/K]<+0.75$ . APOGEE does provide NLTE corrections for Mg, K, and Ca \citep{osorio20}. Unfortunately, APOGEE does not measure Zn and provides very few Cu abundance measurements. Thus, the smoking gun of PISN pollution is missing for this interesting star. Therefore, we applied for an ESO-DDT proposal of one hour to try to accurately measure both \textit{killing} elements, Zn and Cu.

\section{Observations and Analysis}\label{sec:observations}

\subsection{UVES observations}\label{sec:uves}
Our target, 2M13593064+3241036, was observed with UVES at the 8.2\,m VLT Kueyen Telescope in a single observing block (OB) of one hour in service mode, during the night of the 28th of March 2022, under program ID~108.23N5.001. 
A $1\farcs2$ slit was used with $1\times1$ binning in grey sky conditions and an airmass of $\sim2.0$. The seeing was $1\farcs2$ after corrected by airmass. Our settings used dichroic $\#Dic1\,(390+564)$ and provided a spectral coverage between 330 and 660~{nm}.  We corrected each spectrum for the barycentric velocity. The signal-to-noise ratio (SNR) per pixel in the spectra was $\sim$19 at 390~nm, 48 at 510~nm, and 55 at 660~{nm}. The resolving power for this set up is R$\sim45,000$ for the blue part of the spectrum ($330-452$~{nm}) and R$\sim41,500$ for the red ($480-680$~{nm}). The data were reduced using the REFLEX environment \citep{reflex} within the ESO Common Pipeline Library.
\subsection{Stellar Parameters}\label{sec:parameters}
APOGEE analysis pipeline \citep[ASPCAP;][]{aspcap16} used spectral profile fitting to derive for 2M13593064+3241036: \teff=5106\,K, \logg=2.60, and \feh$=-1.79$ \citep{apogee17}. We perform a similar analysis for our UVES optical spectrum by fitting the data with the {\tt FERRE}. {\tt FERRE} is able to match observations with a library of stellar models by interpolating between the nodes of the grid for different parameters. The synthetic models were computed originally by \citet{Aguado2021a}, covering the following range of parameters:
 \begin{itemize}
\item $4500~\mathrm{K} < \teff < 7000~\mathrm{K}$, $\Delta \teff = 250~\mathrm{K}$
\item $1.0  < \logg < 5.0$, $\Delta \logg = 0.5$
\item $-4.0  < \feh < +1.0$, $\Delta \feh = 0.5$
\item $-1.0  < \cfe < +3.0$, $\Delta \cfe = 1.0$
\end{itemize}
The models were computed in one dimension local thermodynamical equilibrium (1D-LTE) with the {\tt ASSET} code and the ATLAS12 stellar atmospheres \citep{sbordone7, kur05}. The atomic and molecular data were taken from Kurucz webpage\footnote{http://kurucz.harvard.edu/} and enriched with the literature as explained in \citep{alle18}. For the analysis we first smoothed and resampled the models to the UVES resolution and performed a running-mean normalization with a 150-pixel window. Then we also normalised the data accordingly and launched the code with the Nash's truncate Newton search algorithm \citep{NASH90}, and cubic order interpolation. Thus we derived for 2M13593064+3241036: \teff$=5036\pm105$\,K, \logg$=2.59\pm0.20$, \feh$=-2.29\pm0.10$, and \cfe$=+0.05\pm0.15$. Atmospheric parameters are in excellent agreement with those derived from the H-band, however, the measured [Fe/H] is 0.5\,dex lower than the ASPCAP value. This discrepancy is certainly unusual and cannot be easily explained since the APOGEE spectrum is of high quality (SNR$\sim$130). We re-analysed the APOGEE spectrum with consistent models from \citet{alle18} and found compatible metallicity $\rm [Fe/H]\sim-1.9$. Therefore, we attribute the discrepancy to a problem that could be related with a problem to the sky or background subtraction in the APOGEE spectrum. Additionally, we found {\tt persisting\_high} flag (residual signal in the detector) within ASPCAP documentation that, in principle, giving the brightness of this object should not be a problem. Yet, we argue that it could possibly play a role and contribute to explain such a difference. Unfortunately, the impact of this metallicity discrepancy significantly influences the abundances, [X/Fe], derived by APOGEE for this star, as is shown in the following Sec. \ref{sec:elemental}. Thus, for sake of caution we will not use APOGEE abundances for this object.

As sanity check, we compared our stellar parameters from spectroscopy with the ones we derived with the \textit{Gaia} colours \citep{gaiaidr3}, applying the calibration by \citet{mucciarelli21} and assuming $\rm [Fe/H]=-2.3$, \teff$=5144\pm100$\,K and \logg$=2.5\pm0.2$. Both set of parameters are in good agreement also with the APOGEE ones. Furthermore, we note that in different visits  APOGEE finds different radial velocities:  MJD\,56389: -15.6\\,km$\,$s$^{-1}$; MJD\,56405: -9.3\,km$\,$s$^{-1}$; and MJD 56412 -8.8\,km$\,$s$^{-1}$ with S/N > 30 for all visits. In addition, \textit{Gaia} RVS \citep{gaia_rvs} is giving  $v_{rad}=-13.5\pm 3.2\rm\,km\,s^{-1}$ with 38 transits. Finally, from the UVES spectrum we derive $v_{rad}=-15.8\pm 1.2\rm\,km\,s^{-1}$. Therefore we cannot exclude that this star is a binary system.

\begin{figure}
\begin{center}
{\includegraphics[width=65 mm, angle=90,trim={ 0.6cm 0.cm 0.cm 2.6cm},clip]{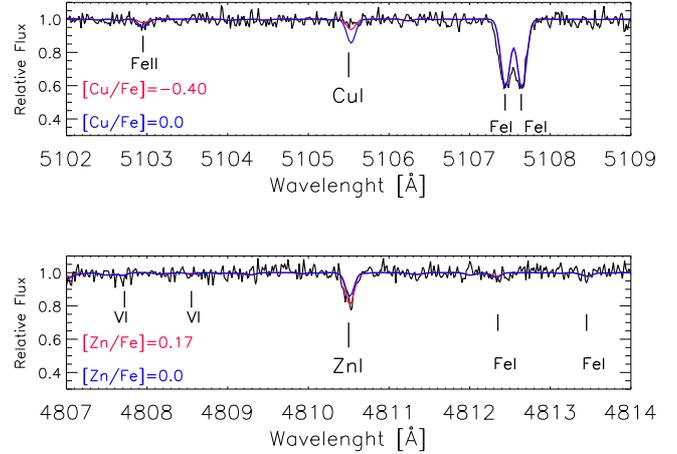}}
\end{center}
\caption{UVES spectrum (black lines) of the star 2M13593064+3241036,  around the absorption lines: \ion{Cu}{I} at 5105\,{\AA} (top); and \ion{Zn}{I} at 4810\,{\AA} (bottom). Two models at different elemental abundances are also shown (blue and red lines). Other metallic absorption lines in the spectra are also labelled.}
\label{fig:uves}
\end{figure}
\input{plots/uves_table.tex}

\subsection{Elemental abundances}\label{sec:elemental}
Detailed chemical abundances analysis has also been performed taking advantage of the {\tt FERRE} code. First, we identified all the resolved lines present in the UVES spectrum and measure their central wavelength by using {\tt splot} routine within  {\tt IRAF}\footnote{IRAF is distributed by the National Optical Astronomy Observatory, which is operated by the Association of Universities for Research in Astronomy (AURA) under cooperative agreement with the National Science Foundation} \citep{tod93}. Secondly, we built for each chemical species a {\tt FERRE}-readable mask that includes all the observed lines. These spectral windows include the vast majority of the information of each elemental abundance. Thirdly, we fixed the stellar parameters to the values derived in the Sec.~\ref{sec:parameters} and launched  {\tt FERRE} again leaving only as a free parameter the chemical abundances. The code calculates and averages the abundance of each element and gives the associated uncertainty. The result of this analysis is presented in Table \ref{results}, and summarised in the following subsections.

\subsubsection{CNO-elements}
The carbon abundance is derived, as explained in Sec.~\ref{sec:parameters}, by using specific models with C as a free parameter. The CH features are clear, resolved, and spread all over the UVES spectrum, especially around the G-band at 4385\,{\AA}. The obtained abundance ratio is almost solar $\rm [C/Fe]=+0.05\pm0.11$, and no significant correction is needed to account for the star's evolutionary stage (0.01\,dex according to \citealt{pla14}). However, for nitrogen we are only able to provide an upper limit from the CN molecular band at 3885\,{\AA}. Finally, since we do not have access to the oxygen triplet at $\sim7775$\,{\AA} we directly take the APOGEE value from the H-band. Taking into account the metallicity from iron lines was off in the ASPCAP calculation we corrected the oxygen ratio accordingly, from $\rm [O/Fe]_{apogee}=0.57$ to $\rm [O/Fe]_{assumed}=1.10$. Unfortunately, the forbidden oxygen lines at 6300 and 6363\,\AA~are dominated by strong sky lines in the UVES spectra so we could not check this high oxygen value. Therefore, we consider  $\rm [O/Fe]_{assumed}$ only tentatively and not used in the following analysis.

\subsubsection{Odd elements: Na, Al, Sc, and K}
Absorption lines from odd elements are well resolved in the UVES spectrum, and {\tt FERRE} provides a good fit for Na (2 lines), Al (2 lines), and Sc (3 lines). We detect several features from interstellar Na around $\sim5985$\,{\AA}. Luckily, they are well separated from the stellar component and we are able to provide an accurate [Na/Fe] value. The strongest K lines are outside of the range of the available UVES spectrum, so we use the corrected value from the H-band in APOGEE, $\rm [K/Fe]_{apogee}=-0.24$ to $\rm [K/Fe]_{assumed}=0.24$. We consider this value cautiously for a number of reasons: a)~K lines in the H-band are weak in this metallicity regime; b)~K is derived by ASPCAP with the metallicity as a free parameter which could potentially introduce some deviation since we know that [Fe/H] was overestimated. Therefore, as explained in Sec. \ref{sec:uves} we do not use K in our analysis (i.e. we give weight equal zero).

\subsubsection{Alpha-elements: Mg, Si, S, Ca, and Ti}
The $\alpha-$elements show strong absorptions, even in very metal-poor stars. In particular, we detect 9, 1, 14, and 20 lines for Mg, Si, Ca, and Ti, respectively. We do not use the saturated \ion{Ca}{ii} lines at 3933\,{\AA} and 3968\,{\AA} since the size of the spectral window would be gigantic and many other lines actually do live in their wings. We included two ionised states of titanium, \ion{Ti}{i} and \ion{Ti}{ii}, with 4 and 16 detected lines respectively. The measured abundances of the two Ti species agreed with in errors, and the assumed value in our fits is the average of the two (See Table \ref{results}).  Finally, we discard the infrared sulphur measurement from APOGEE, plotted in Fig. \ref{fig:example}, since no appropriate line is available in the APOGEE spectrum and no strong absorption lines populate the optical spectrum. 

\subsubsection{Iron peak-elements: V, Cr, Mn, Fe, Co, and Ni}
The relatively high SNR and resolution of the UVES spectrum allowed us to derive iron peak-elements with high accuracy (number of lines): V (8), Cr (6), Mn (3), Co (6), and Ni 
(5). For \ion{Fe}{i} we identified up to 231 lines with a mean metallicity of  $\rm [\ion{Fe}{i}/H]=-2.31$, and we adopted this as the metallicity of 2M13593064+3241036. On the other hand, we obtained $\rm[\ion{Fe}{ii}/H]=-2.38$ from 7 isolated lines.

\subsubsection{Killing elements: Cu and Zn}
As previously explained, Cu and Zn are key elements for our PISN analysis. Thus the required SNR in the UVES spectrum was based on these two elements, so that a minimum detection threshold of $\rm [Cu/Fe]=-0.40$ and $\rm [Zn/Fe]=-0.70$ could be potentially detected from the \ion{Cu}{i}  (5105\,{\AA}) and \ion{Zn}{i} (4810~{\AA}) lines. In Fig.~\ref{fig:uves} the targeted lines for Cu and Zn, and their derived values are shown.

\subsubsection{Neutron-capture elements: Sr, Y, Zr, Ba, and Eu}
 We also derived ionised species for neutron-capture elements from the UVES spectrum. We detect several of the typical s-process elements (number of lines): Sr (2), Y (4), Zr (3), and Ba (3). Additionally, corresponding to a r-process production, the strongest Eu lines (4129\,{\AA} and 4205\,{\AA}) were detected and measured. 
 
\subsection{NLTE corrections}
 We note that the way we derived elemental abundances with spectral windows does not allow direct NLTE corrections since we are fitting several lines at the same time. However, we estimated an overall NLTE correction by averaging the individual corrections over the relevant lines. We employed NLTE corrections provided in the literature:
 \begin{itemize}

 \item \citet{mashonkina2016}\footnote{http://spectrum.inasan.ru/nLTE/} to calculate both \ion{Fe}{i} and \ion{Ti}{ii}. The average \ion{Fe}{i} NLTE correction is 0.04\,dex and the dispersion line to line is also low with the majority of the lines between 0.02-0.05\,dex. \ion{Ti}{ii} corrections go in the other way but are also small with an average of $-$0.03\,dex.

 \item (b)\,\citet{lind12}\footnote{http://www.inspect-stars.com/} for Na. The lines/individual corrections are 5889\,{\AA}/$-$0.42\,dex and 5895\,{\AA}/$-$0.45\,dex, giving a final value of 0.44\,dex. These are, as expected, the largest in this giant star.

 \item \citet{Mashonkina2007, Bergemann2013, Bergemann2017}\footnote{https://nlte.mpia.de/gui-siuAC\_secE.php} for \ion{Ca}{i}, \ion{Si}{i}, and \ion{Mg}{i}, respectively. \ion{Ca}{i} are comparatively large (0.13\,dex), while \ion{Si}{i}, and \ion{Mg}{i} are small and at the level of $-$0.02 and 0.04, respectively. This is not surprising for a very metal-poor K type star \citep[e.g.][]{alexeeva18}.
 
 \item We also applied a $+$0.30\,dex correction for Al based on NLTE calculations by \citet{nordlander17}.
  \end{itemize}


\section{PISN Descendants Candidates}\label{sec:validation}
\begin{figure*}
  \centering

     \begin{subfigure}[b]{0.39\linewidth}
         \centering
         \includegraphics[width=0.71\hsize,angle=90]{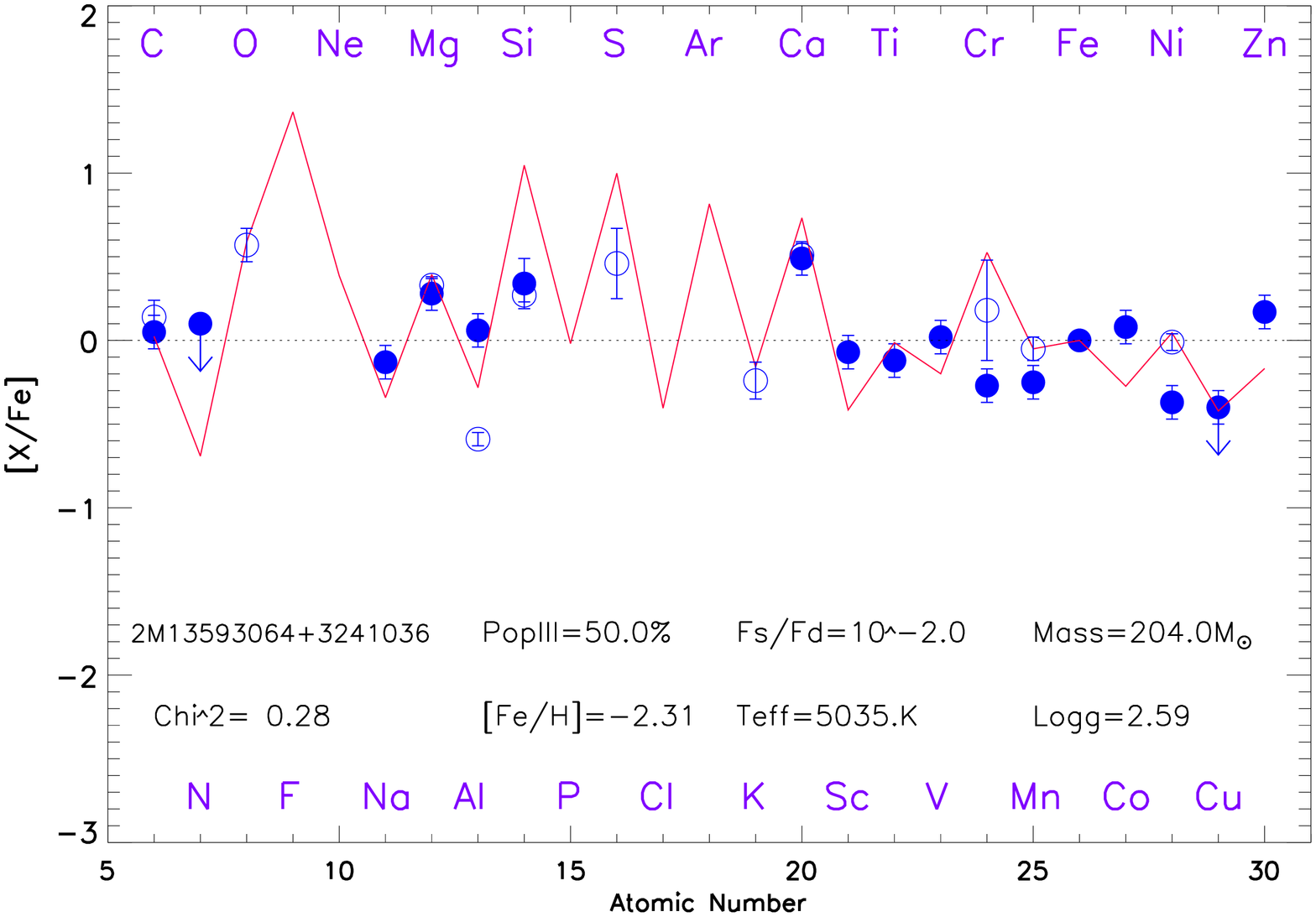}
         \caption{2M13593064+3241036}
         \label{fig:ddt}
     \end{subfigure}
     \begin{subfigure}[b]{0.39\linewidth}
         \centering
         \includegraphics[width=0.71\hsize, angle=90]{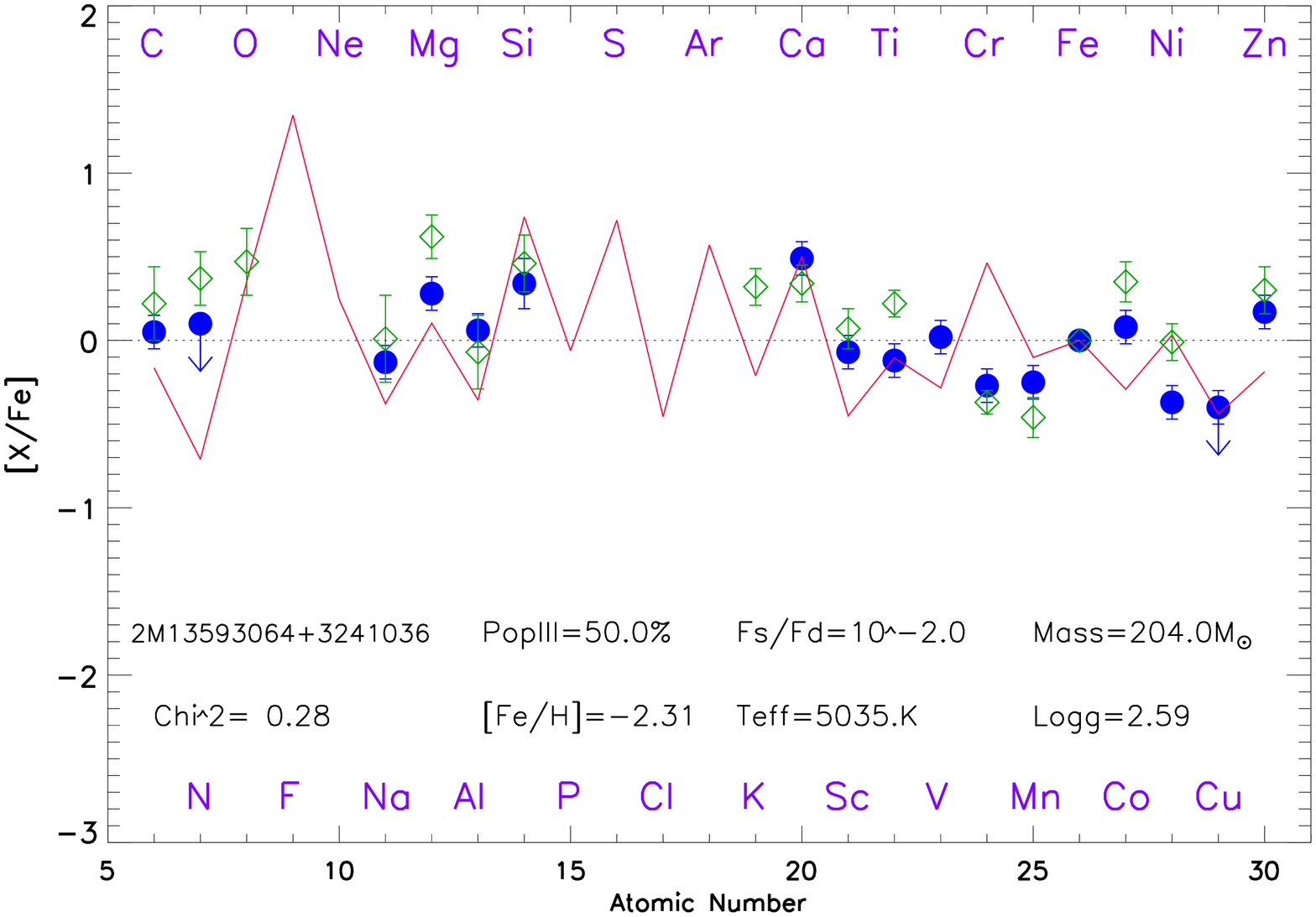}
         \caption{Halo abundance pattern}
         \label{fig:cayrel}
     \end{subfigure}
     
      \begin{subfigure}[b]{0.39\linewidth}
         \centering
         \includegraphics[width=0.71\hsize,angle=90]{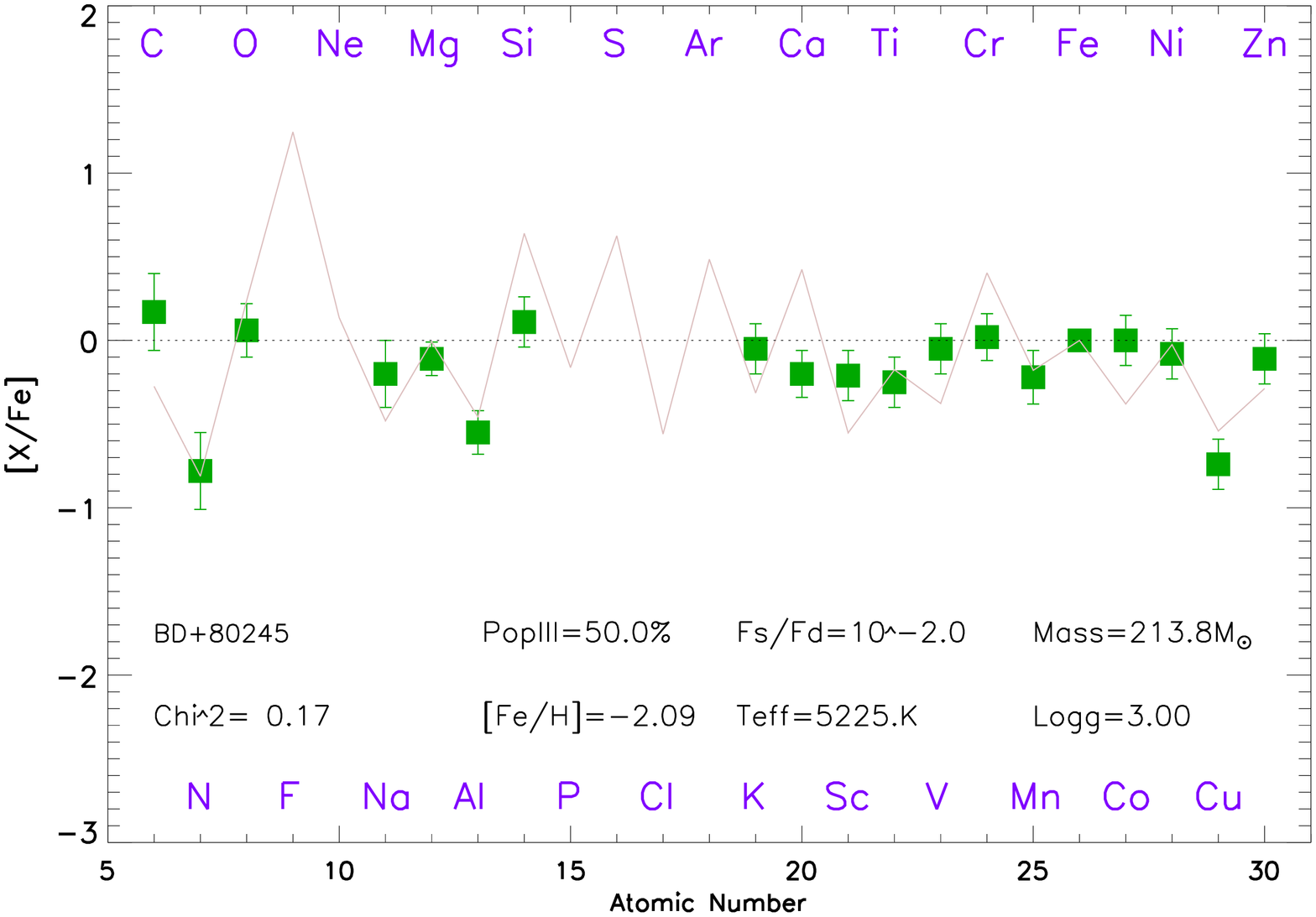}
         \caption{BD+80$^{\circ}$245}
         \label{fig:salva}

     \end{subfigure}
     \begin{subfigure}[b]{0.39\linewidth}
         \centering
         \includegraphics[width=0.71\hsize,angle=90]{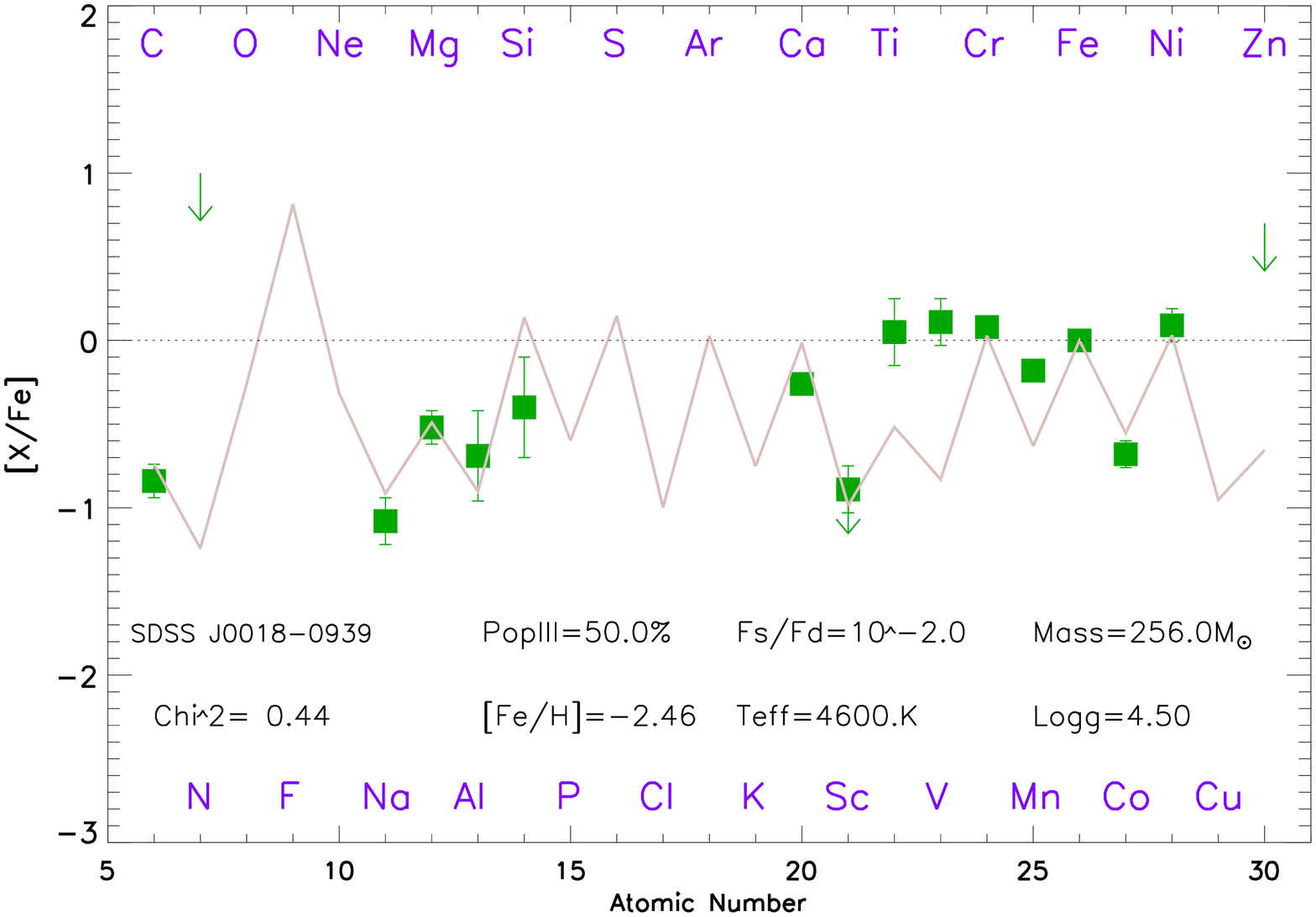}
         \caption{SDSS\,J0018$-$0939}
         \label{fig:aoki}

     \end{subfigure}
         \begin{subfigure}[b]{0.39\linewidth}
         \centering
         \includegraphics[width=0.71\hsize,angle=90]{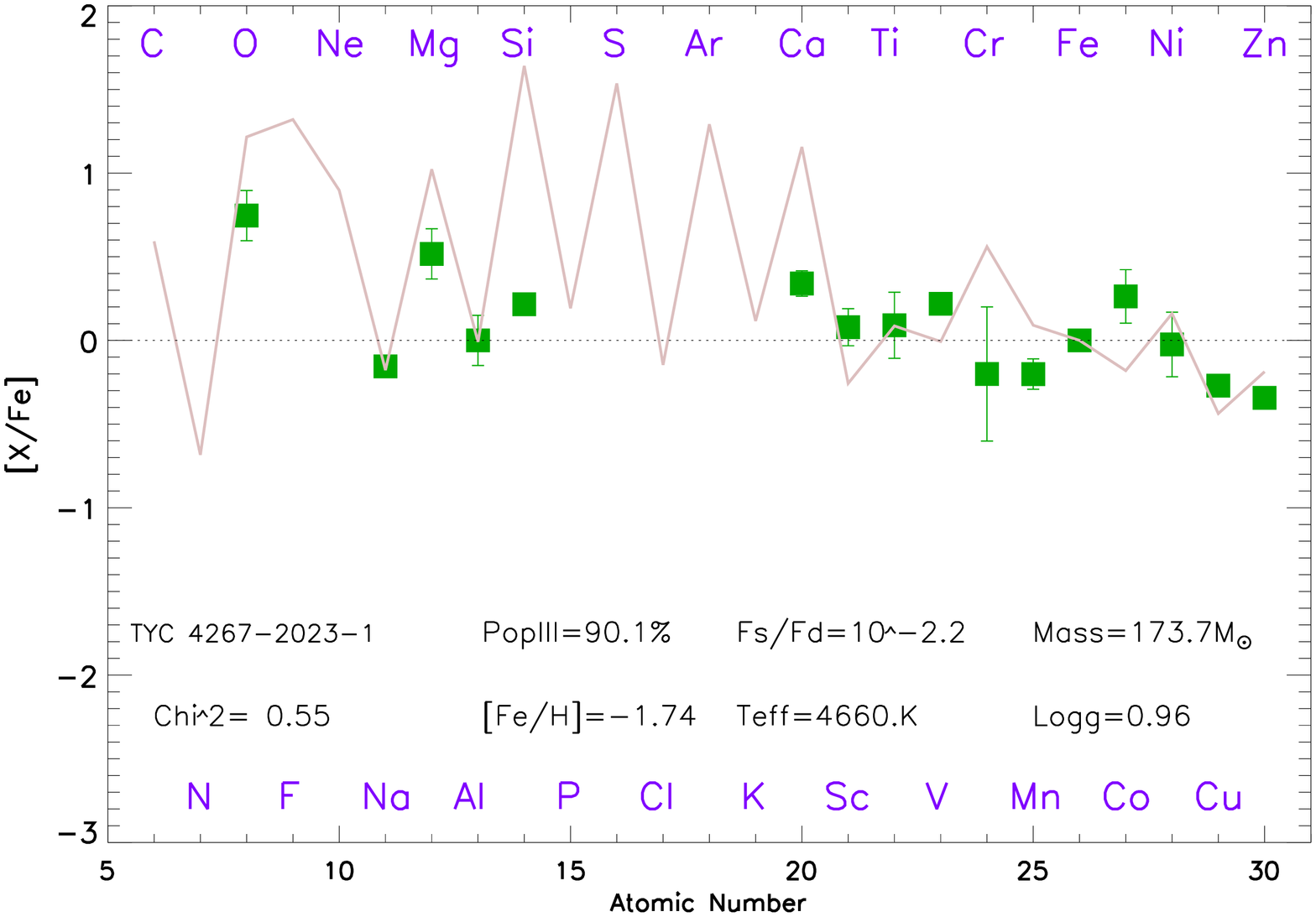}
         \caption{TYC\,4267-2023-1}
         \label{fig:mince1}

     \end{subfigure}
     \begin{subfigure}[b]{0.39\linewidth}
         \centering
         \includegraphics[width=0.71\hsize,angle=90]{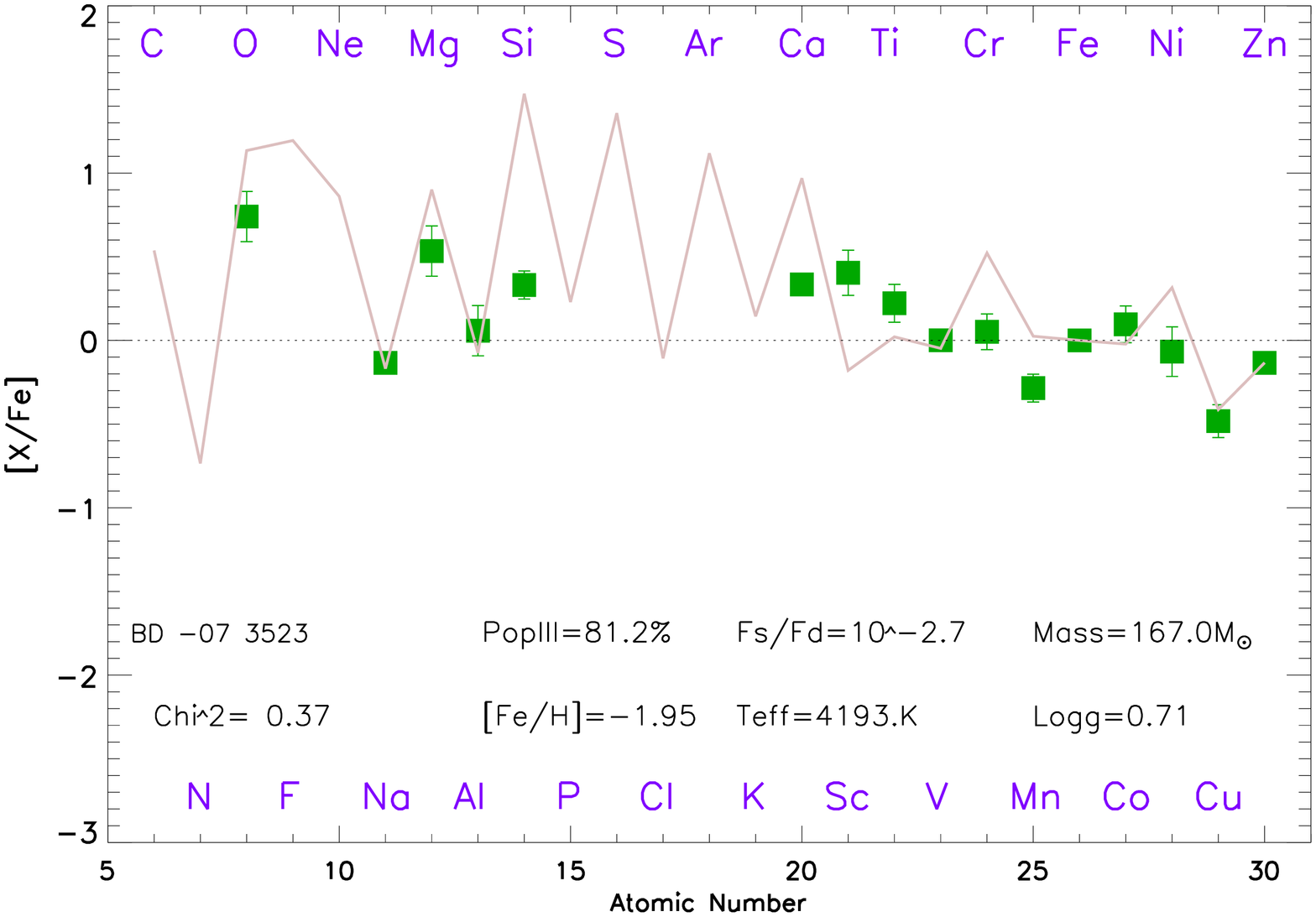}
         \caption{BD\,$-$07\,3523}
         \label{fig:mince2}

     \end{subfigure}
 
         \begin{subfigure}[b]{0.39\linewidth}
         \centering
         \includegraphics[width=0.71\hsize,angle=90]{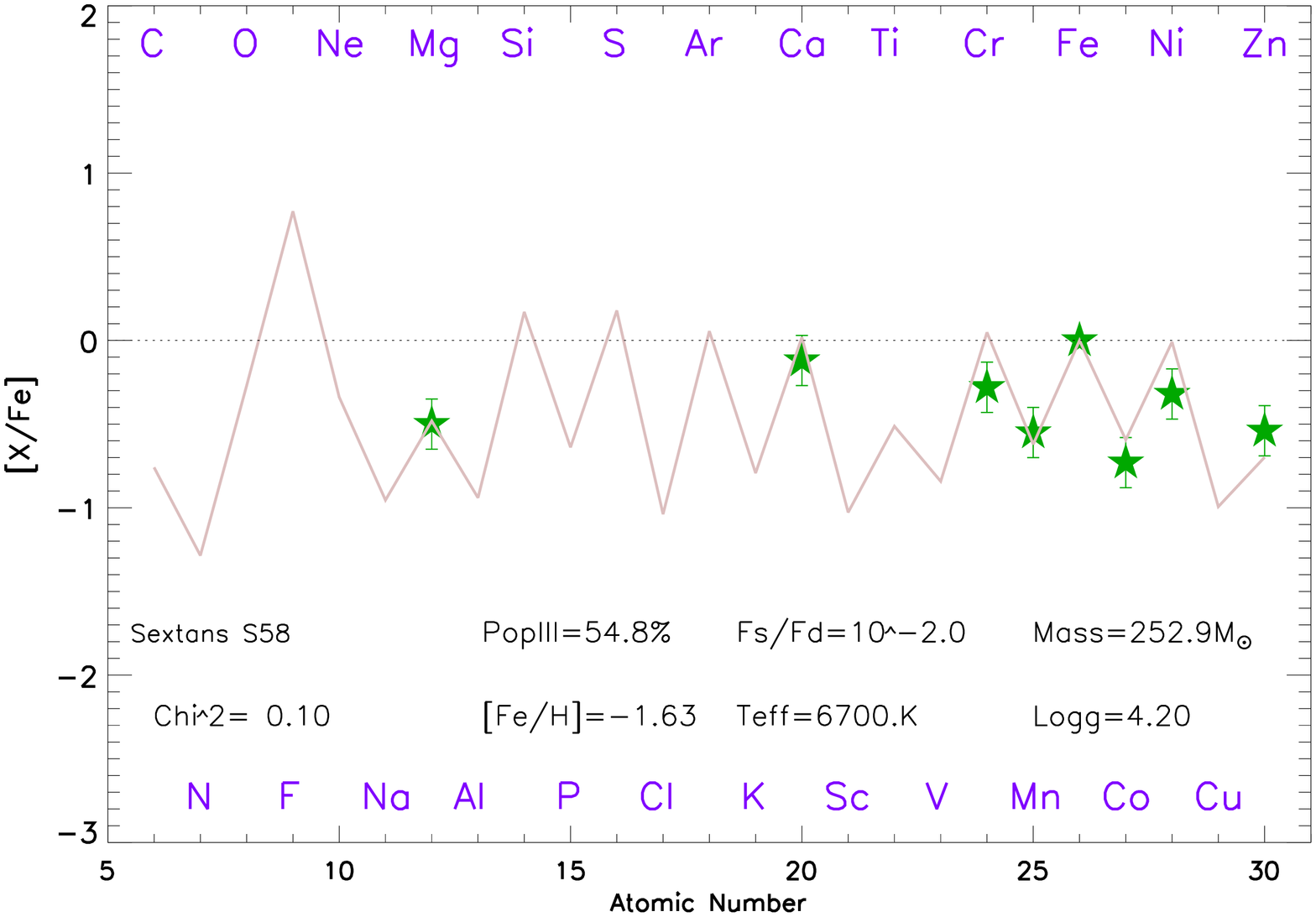}
         \caption{Sextans\,S58}
         \label{fig:sextans}

     \end{subfigure}
     \begin{subfigure}[b]{0.39\linewidth}
         \centering
         \includegraphics[width=0.71\hsize,angle=90]{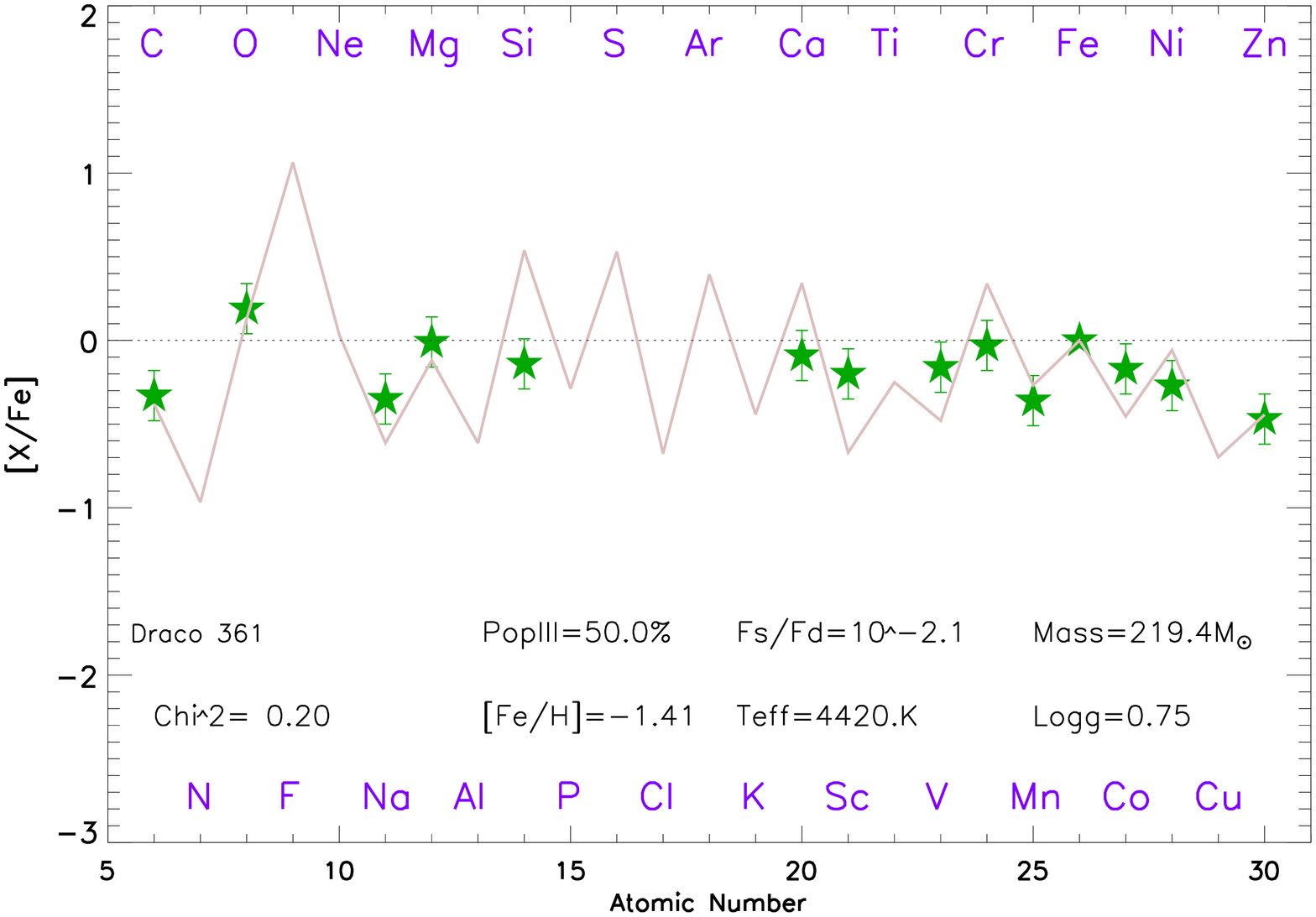}
         \caption{Draco\,361}
         \label{fig:draco}
     \end{subfigure}
  \caption{Results of the {\tt PISN-explorer} analysis. Points are measured chemical abundances while lines are best fits. Listed on the plot are parameters of the fit, along with \teff, $\log g$, and [Fe/H].
{\it Top row:} Our new analysis is presented (blue filled circles) and compared to the APOGEE results (a, blue open circles); and the average abundance pattern and the best fit of the \citet{cay04} sample (b, green open diamonds). {\it Second row:} Previously published candidates for dominant PISN enrichment (green squares), SDSS J0018$-$0939 (c; \citealt{aoki14}) and BD+80$^{\circ}$245 (d; \citealt{salvadori19}). {\it Third row:} Candidates selected from the MINCE survey (e and f; \citealt{mince}). {\it Bottom row:} Promising candidates for PISN descendants (green star symbols) in the Sextans (g; \citealt{shetrone2001}) and Draco (h; \citealt{cohen09}) dwarf spheroidal galaxies.} 
\end{figure*}

\subsection{Our observed candidate: 2M13593064+3241036}\label{ddt}
In Sec. \ref{sec:methodology} we reanalysed with {\tt FERRE} the chemical signature of our PISN candidate.  As explained in Sec. \ref{sec:parameters} the fact that 2M13593064+3241036 is more metal-poor than previously published with APOGEE has a direct impact on the analysis and makes the \textit{killing} ratios [Cu/Fe] and [Zn/Fe] higher than expected, with catastrophic consequences. Additionally, we consider the K measurement as tentative with lower weight in the analysis (see Sec. \ref{sec:elemental}). Furthermore, the [Mg/Al] ratio has dramatically increased by $0.69$\,dex which clearly attenuates the odd-even effect. Still some key features remain, such as high [Si/Fe] value, and $\rm [Ca/Mg]>0$, but the importance of those is relatively smaller when identifying PISN descendants.

In Fig. \ref{fig:ddt} we show the elemental abundances from Table~\ref{sec:uves} together with the best {\tt PISN-explorer} fit and parameters. The best fit gives significantly different results than the original analysis: $m_\text{PISN}=204.0\,\msun$ (prev.~$193.8\,\msun$), and $f_\text{PISN}=50\%$ (prev. 66\%), and $f_*/f_{\rm dil}=10^{-2.1}$ (prev.~$10^{-2}$). The quality of the fit is now worse (See Fig. \ref{fig:example}), in particular the killing element Zn is not anymore well reproduced by the models. Additionally, the odd-even effect is attenuated and the best fit (red line) is unable to reproduce it which is far from we could expect from a PISN descendant. That is consistent with the fact that now the best fit is giving a $f_\text{PISN}$ value that is in the limit of the grid, which could be an indication of even lower PISN contribution.
Three key results therefore forced us to be cautious on reaching strong conclusions on whether 2M13593064+3241036 is a PISN descendant: (a)~the quality of the fit has significantly decreased with the analysis of the higher quality UVES spectrum; (b)~the odd-even effect is now clearly reduced; and (c)~the high Zn value does not allow us to clearly identify the smoking gun of PISN production. The candidate 2M13593064+3241036 could therefore still be a regular halo star. The following is an additional test we propose to verify this hypothesis.

The authors in \citet{Cayrel04} derived precise elemental abundances for 35 very metal-poor halo stars, considered chemically unmixed \citep{spi05}. These abundances adequately averaged and NLTE corrected by \citet{andrievsky2010} can be understood as the \textit{mean chemical signature} for a regular (C-normal) giant halo star polluted almost exclusively by CCSN. This ``representative'' abundance pattern is not polluted by PISN at the 99.9\% level, and is therefore useful as a comparison sample to test our methodology. Thus, following the methodology described in Sec. \ref{sec:methodology}, we analyzed this average abundance pattern for a regular giant halo star having as result $f_{\rm PISN}=50$\%; $f_*/f_{\rm dil}=10^{-2.0}$; and $m_{\rm PISN}=231\msun$. In Fig. \ref{fig:cayrel} we show the comparison between 2M13593064+3241036 and the regular giant halo star pattern. As in the case of our candidate we obtained the minimum value of PISN contribution  while the mass of the progenitor is slightly higher. In fact, both objects and their best fits seem similar. Main differences are concentrated in lighter elements such us the CNO family where the regular halo star pattern shows smother trend. In addition to it, the odd-even effect is even less clear in 2M13593064+3241036 in the Na-Mg-Al-Si range. Furthermore, the super solar Zn abundance in the regular halo star pattern (as in our APOGEE candidate) is a strong indication of no PISN contribution. The super-solar [Zn/Fe] makes it difficult to establish an unambiguous difference between this star and the average halo star. Therefore we could conclude that 2M13593064+3241036 is likely like other halo stars with no major contribution from PISN pollution.

\subsection{Previously reported candidates}
In the literature, two stars have been reported as being probable descendants of PISNe: BD+80$^{\circ}$245 \citep{salvadori19}, and SDSS\,J0018$-$0939 \citep{aoki14}. To verify our methods, and compare our approach to their published results, we also analysed these stars with the {\tt PISN-explorer}, and the result is shown in Figs~\ref{fig:aoki} and \ref{fig:salva}.

The star BD+80$^{\circ}$245 \citep{carney97, ful10, ivans03, roe14}, is a low-$\alpha$ halo star (Fig.~\ref{fig:salva}), which was initially proposed as a PISN descendant \citep{salvadori19} based on the low abundance of the \textit{killing} elements (Cu and Zn). Their originally derived PISN parameters were $f_{\rm PISN}=50$\%; $f_*/f_{\rm dil}=10^{-4.0}$; and $m_{\rm PISN}=223\msun$. Our analysis, as explained, is based on the same set of models \citep{heger2002, salvadori19} but the way we fit the data with {\tt FERRE} is different (see Sec. \ref{sec:methodology}). The best solution we get is $f_{\rm PISN}=50$\%; $f_*/f_{\rm dil}=10^{-2.0}$; and $m_{\rm PISN}=213.8\msun$ with a $\chi^2=0.17$, which is in good agreement with \citet{salvadori19}. Interestingly, the $f_*/f_{\rm dil}$ factor is significantly different in the two analysis while the quality of the fit is similar. The reason for that is that $f_*/f_{\rm dil}$ is the least sensitive parameter (see Sec.~\ref{sec:models}) within the models and the {\tt FERRE} code tends to go to the limit of the grid.

SDSS\,J0018$-$0939 is another interesting star that was proposed to be a PISN descendant by \citet{aoki14}. This star has remarkably low $\alpha$-abundances ($\rm [Mg/Fe]=-0.52$, $\rm [Ca/Fe]=-0.26$), see Fig.~\ref{fig:aoki}. The authors excluded that the peculiar chemical pattern could not be produced by CCSN or SN\,Ia. They concluded that the most likely origin is PISN and they attribute a $f_{\rm PISN}=100$\% value. Unfortunately, no informative upper limits for Cu and Zn were measured in this star. We used the {\tt PISN-explorer} to derive PISN parameters from the measured abundances by \citet{aoki14} and found $f_{\rm PISN}=50$\%; $f_*/f_{\rm dil}=10^{-2.0}$; and $m_{\rm PISN}=256\msun$ with $\chi^2=0.44$. Although the $\chi^2$ is elevated, we are able to reproduce the chemical signature of lighter elements (See Fig. \ref{fig:aoki}) but the models failed for iron-peak elements. The low $\alpha$ ratios lead to high mass of the progenitor ($256\msun$) in agreement to what was claimed by \citet{aoki14}. However, the $f_{\rm PISN}=50$\% value suggests that this star could be polluted by PISN but a higher contribution should be attributed to normal Pop\,II stars exploding as CCSN. Further observations in order to detect and measure \textit{killing} elements is highly required to confirm the percentage of PISN contamination we see in this star.

 The \textit {Pristine} survey \citep{sta17I} derives metallicities for halo stars based on narrow band filter photometry. In this context, there is also a remarkable observational effort  to derive Cu and Zn in metal-poor stars \citep{caffau22}. The authors found, among others, three interesting candidates to be polluted by PISN production using also the {\tt PISN-explorer}. In particular, in their Fig. 12, they show the chemical pattern of TYC\,1118-595-1,  a very metal-poor star with $\rm [Fe/H]=-2.12$, and derived $f_{\rm PISN}=50$\%; $f_*/f_{\rm dil}=10^{-2.3}$; and $m_{\rm PISN}=193\msun$. TYC\,2207–992–1 and TYC\,1194–507–1 are also reported together with a best fit that suggested that they could be enriched by PISN up to $f_{\rm PISN}=83$\% and 90\%, respectively. However, the probability of such is significantly smaller than for TYC\,1118-595-1 due to the lower quality of the fit. According to what is explained in Sec. \ref{sec:methodology}, we consider that TYC\,1118-595-1 is likely reflecting the theoretical predictions in their chemical pattern in a percentage that could be to the order of $f_{\rm PISN}=50$\% or slightly smaller. TYC\,2207–992–1 is also a very promising candidate with a progenitor mass $m_{\rm PISN}=170\msun$ while TYC\,1194–507–1 has more uncertain origin due to the low quality of the fit. 

 We also mined the bibliography including high-resolution analysis of halo stars looking for interesting candidates. We found that the vast majority of interesting stars published before 2018 are already included in the JINA database. However, \citet{xing2019} found an interesting star, J1124+4535, with $\rm [Fe/H]=-1.27$ and $\rm [Zn/Fe]=-0.37$. We also analyzed its chemical signature with the {\tt PISN-explorer} and found $f_{\rm PISN}=50$\%; $f_*/f_{\rm dil}=10^{-2.4}$; and $m_{\rm PISN}=231\msun$. The quality of the fit is relatively good ($\chi^2=0.29$) but the absence  of Cu measurement or upper limit prevent us to conclude this object is polluted by PISN. We propose to re-observe this interesting object with peculiar chemistry to confirm its origin.

\subsection{PISN candidates in classical dwarf galaxies}
It is commonly accepted that the classical dwarf spheroidal galaxies are massive enough to retain the chemical products of PISN explosions \citep[see e.g.][]{brom03, salvadori08}. However, their total number of stars is
much lower compared to more massive systems. Consequently, the expected fraction of PISN descendants in classical dwarf galaxies is expected to be significantly higher than in the Milky Way.
Therefore, classical dwarf galaxies are interesting places to search for PISN descendants. Luckily, within the JINA database there are $\sim50$ stars from Fornax, Sextans, Draco, and other dwarf satellites. In Figs.  \ref{fig:sextans} and \ref{fig:draco} two interesting examples of the {\tt FERRE} analysis are shown, Sextans\,S58 \citep{shetrone2001} and Draco\,361 \citep{cohen09}. The quality of both fits are remarkably high with  $\chi^2=0.10,0.20$, respectively. Additionally, the \textit{killing} element Zn, is largely subsolar in the atmosphere of the two stars $\rm [Zn/Fe]\sim-0.45$. Actually, the case of Sextans\,S58 with $f_{\rm PISN}=55$\% suggests that the majority of the material this star formed from was polluted mostly by PISN with a mass of $m_{\rm PISN}=253\,\msun$. 

Finally, we also identified another star from JINA database, Draco\,3053 \citep{cohen09}. This star is also Zn-poor with $\rm [Zn/Fe]=-0.3$ and the best set of parameters derived with the {\tt PISN-explorer} are $f_{\rm PISN}=50$\%; $f_*/f_{\rm dil}=10^{-2.0}$; and $m_{\rm PISN}=218\msun$; and $\chi^2=0.19$. In this case, the high quality of the fit and the low value of Zn suggest that this star could be indeed polluted by PISN in a percentage close to $f_{\rm PISN}=50$\%. Further observations of all of the three candidates in Draco and Sextans galaxies are required.


 \section{Discussion}
 In the previous sections we have validated and applied our proposed {\tt PISN-explorer} methodology to find PISN descendants. Although the 2M13593064+3241036 candidate seems not to be as interesting as expected (mostly due to the initial overestimation of metallicity in APOGEE), its study was useful to understand the capabilities of our methodology. As shown, its chemical signature, and in particular the \textit{killing} elements Cu and Zn, is not significantly different from that of an average \textit{pattern} for regular giant halo stars. Indeed, when the contribution of Pop\,II stars exploding as normal CCSN becomes $> 50\%$, the peculiar chemical signatures left by different Pop\,III star progenitors are essentially lost (Vanni et al. in prep.) 
 
 A different situation is presented for BD+80$^{\circ}$245, shown in Fig.~\ref{fig:salva}. For this star we confirm the result provided by \citet{salvadori19} and we conclude that it is at least partially (but genuinely) polluted by PISN. The clearly low value of the \textit{killing} elements that this stars shows is indeed the smoking gun of PISN contamination.  
 On the other hand, the analysis of SDSS\,J0018$-$0939 shows that while very different from a regular halo star (see Sec. \ref{ddt}), the percentage of material produced by PISN may not be as high as originally suggested by \citet{aoki14}. In any case, some further high-resolution observations of this interesting star are needed to better constrain the amount of Cu and Zn and shed light over the exact amount of material that effectively come from PISN production. At this point we propose using facilities that will be available in the next generation of 30\,m telescopes. Moreover, the agreement of our analysis with previous works suggests that the {\tt PISN-explorer} is efficient in characterising candidates.


 \begin{figure*}
\begin{center}
{\includegraphics[width=100 mm, angle=90,trim={ 0.cm 0.cm 0.cm .cm},clip]{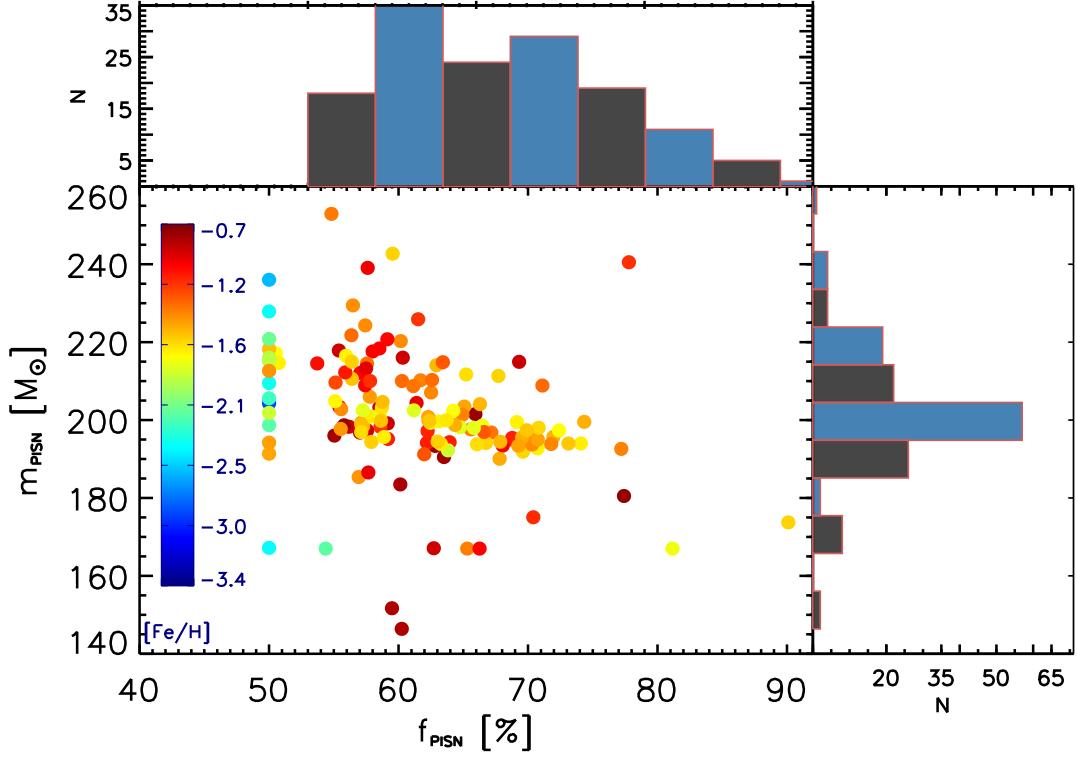}}
\end{center}
\caption{Our catalogue of 166 candidate PISN descendants, shown in the $f_{\rm PISN}-m_{\rm PISN}$ plane, colour-coded by metallicity. Histograms of the candidates are also shown on top and right side panels.}
\label{fig:masspop}
\end{figure*}

 A very challenging issue when identifying the descendants of the very massive first stars is how to discriminate between them and a regular halo star (mostly polluted by Pop\,II exploding as CCSN). By comparing Fig. \ref{fig:cayrel} with \ref{fig:salva}, \ref{fig:aoki}, \ref{fig:sextans}, and \ref{fig:draco} it is clear that the chemical signatures that correspond to good fits in this work ($\chi^{2}\lesssim 0.20$) are very different than the mean abundance pattern for a regular giant halo star \citep{Cayrel04}. The much higher CNO abundances together with high $\alpha$'s (i.e. Mg, Ti) led to a significantly higher $\chi^2$. Obviously, this is deeply related with the fact that the regular halo star pattern is not Zn-poor, which is the most robust indicator. Additionally, as shown in Fig. \ref{fig:cayrel}, the odd-even effect is clearly lost for atomic numbers higher than Z=19, i.e. for most of the iron-peak elements. According to our novel {\tt PISN-explorer}, the most efficient way to separate PISN candidates from other halo field stars is a combination of all the criteria presented in Sec. \ref{sec:selection}, low values of $\chi^2$, and the killing elements deficiency $[\rm Cu,Zn/Fe]<0$. In Figs. \ref{fig:mince1} and \ref{fig:mince2} two interesting examples of promising candidates from MINCE survey with subsolar Cu and Zn abundances are shown.

 Moreover, the {\tt PISN-explorer} allowed us to successfully identify very interesting PISN descendants candidates in classical dwarf galaxies (see Fig.~\ref{fig:mdf}). Note that none of them are found in the less massive ultra-faint dwarf galaxies. These systems, which are smaller and thus have lower binding energy than classical dSph galaxies, are probably not able to retain the chemical products of energetic PISN (Rossi et al. in prep).  Unfortunately, since those from classical dwarfs are typically distant and therefore faint, the next generation of 30\,m class telescopes will be required to derive systematically their \textit{killing} elements. 

A final remark concerns the MDF of our 166 selected candidates shown in Fig. \ref{fig:mdf}. It is indeed extremely interesting to see that its main peak is located at $\rm [Fe/H]=-1.7$, i.e. exactly at the level predicted by the cosmological models for the MW formation of \cite{bennassuti2017} (see their Fig.~9) and close to the values provided by the parametric study of \cite{salvadori19}, $\rm[Fe/H]\approx -2$, and by the inhomogeneous chemical enrichment model by \cite{karlsson2008}, $\rm[Fe/H]\approx -2.5$. This result is quite remarkable since the {\tt PISN-explorer} does not use the iron abundance to select PISN candidates. As explained in Sec.~\ref{sec:selection}, indeed, our selection is solely based on the predicted chemical abundance ratios, [X/Fe]. We should also note that it is very likely that our sample, and thus our selected candidates, are biased towards higher [Fe/H]. In fact, both APOGEE and GALAH rapidly decrease in accuracy in the very metal-poor regime, which is possibly the origin of the second peak at higher [Fe/H]$\approx -1$. Finally, we recall that the iron abundances of PISN descendants is expected to vary in a broad range, $-5<\rm[Fe/H]<0.5$ \citep[see Fig.~7 of][]{salvadori19}. Therefore, it is quite normal that in our MDF we find a second and smaller peak at $\rm [Fe/H]=-3.0$. This peak is indeed made by PISN candidates selected from the JINA database, which is naturally biased towards Fe-poor stars.
 
In Fig. \ref{fig:masspop} we display the $f_{\rm PISN}-m_{\rm PISN}$ distribution for our 166 selected candidates colour coded by metallicity. While the histograms of $f_{\rm PISN}$ are peaked at lower contribution and monotonically decreases, which is expected, the distribution of $m_{\rm PISN}$ is more condensed around $200\,\msun$. We also see a more metal-poor population lying at  $f_{\rm PISN}=50$\%. As explained in Sec. \ref{sec:selection} the vast majority of those values should be avoided when selecting candidates. However, as we pointed out, after visual inspection we could consider them as a candidates whether they show a remarkably good fit and Cu and Zn abundances are subsolar. This population was selected following this approach and some of them (due its visibility and/or special features) are included in our golden sample. However, some others are quite faint and we propose to observe at higher SNR when possible with the future high-resolution spectrographs.

\section{Conclusions}\label{sec:concluisons}
We have presented a new methodology to identify candidates that have been significantly polluted by PISN in the early Universe. We mined different datasets in order to find chemical patterns that match with what is expected to find if significant PISN production took place. We summarise here the main conclusions of this work: 

\begin{itemize}
    \item Current and upcoming large spectroscopic surveys (APOGEE, GALAH, GES, 4MOST, WEAVE), and existing databases such as JINA, are an invaluable tool to unveil the origin and characteristics of the very massive first stars, which exploded as PISN.
    \item The {\tt PISN-explorer}, using the {\tt FERRE} code in combination with theoretical predictions \citep{salvadori19}, is a very efficient methodology when selecting PISN descendant candidates in large databases.
    \item The MDF of the selected candidates is a confirmation of the predicted peak in PISN production at $\rm [Fe/H]\sim-1.7$ \citep{bennassuti2017}.
    \item \textit{killing} elements (Cu and Zn), low values of $\alpha$-elements (Mg and Ti), and clear odd-even effect (e.g.~high [Mg/Na], and/or high [Mg/Al]), are indicators of PISN production. Therefore, they could be used to efficiently select candidates.
    \item It is possible that 2M13593064+3241036 was mostly  polluted by PISN production but the presence of Zn on its atmosphere does not allow us to confirm this hypothesis. BD+80$^{\circ}$245 contains significant material that was formed during a PISN event. The $f_{\rm PISN}=$ value previously reported for SDSS\,J0018$-$0939 (100\%) could be significantly lower according to its chemical signature.
    \item Sextans\,S58 is the most promising candidate and maybe mostly formed out of PISN material. To confirm this hypothesis further high-resolution follow-up is needed to complete the chemical signature already reported in the literature. Draco\,361 and Draco\,3053 are also excellent candidates selected from classical dwarf galaxies. 
    \item From our selected 166 candidates we propose 45 of them as a golden catalogue with the most promising and visible stars for future follow-up. 
\end{itemize}

Future high-resolution facilities mounted in the Extremely Large Telescope such as ANDES \citep{marconi22} and other facilities will provide and unbeatable opportunity to observe stars in dwarf satellites, where it is more likely to find PISN descendants. In addition to it, the next generation of large spectroscopic surveys such as WEAVE (Jin et al., in press) and 4MOST \citep{deJong2019} with high resolution observations will be an unbeatable place to test the {\tt PISN-explorer} and finally finding the descendants of the very massive first stars.

\section{Data Availability and online material}
All the spectroscopic data reduced and analyzed for the present article are fully available under request to the corresponding author\footnote{david.aguado@unifi.it}. A table with the detected lines in the UVES spectrum of 2M13593064+3241036 is included as online material. 

\section*{Acknowledgements}
We thanks the anonymous referee for positive and constructive comments. The authors of this work really thank Carlos Allende Prieto (Instituto de Astro\'isica de Canarias) for insightful discussion about the capabilities of the {\tt FERRE} code. We warmly thank all the members of the NEFERTITI group of the University of Florence for insightful discussions. We also thank the \textit{streams} group at the University of Cambridge for fruitful interactions.
These results are based on VLT/UVES observations collected at the European Organisation for Astronomical Research (ESO) in the Southern Hemisphere under program ESO ID 108.23N5.001. This project has received funding from the European Research Council (ERC) under the European Union’s Horizon 2020 research and innovation program (grant agreement No. 804240). DA, SS, AS, IV, VG and IK acknowledge support from the European Research Council (ERC) Starting Grant NEFERTITI H2020/808240. SS acknowledges support from the PRIN-MIUR2017, prot. n. 2017T4ARJ5. EC and PB gratefully acknowledge support from the French National Research Agency (ANR) funded project ``Pristine'' (ANR-18-CE31-0017). AMA acknowledges support from the Swedish Research Council (VR 2020-03940).




\bibliographystyle{mnras}

\bibliography{biblio}


\end{document}

%% file: plots/uves_table.tex
\begin{table}
\renewcommand{\tabcolsep}{2pt}
\caption{Stellar parameters, abundances, ratios, errors and number of detected lines for individual species derived in 2M13593064+3241036 from UVES and APOGEE data. \label{results}}
\resizebox{1.\linewidth}{!}{
\begin{tabular}{ccccccc}
\multicolumn{7}{l}{\textit{Gaia} DR2 id 1457695618046140800}\\
 \multicolumn{7}{l}{  $\rm RA(deg)=209.877547$, $\rm DEC(deg)=32.684315$}\\
\multicolumn{7}{l}{  $v_{rad}=-13.45$\,km$\,$s$^{-1}$, $\teff=5035$\,K, \logg=2.59} \\
 \hline
Species & $\log\epsilon$\,(X)$_{\odot}^1$ & $\log\epsilon$\,(X) & $\mbox{[X/Fe]}^{2}$ & $\sigma_{\mbox{[X/Fe]}}$ & $N$& $\mbox{[X/Fe]}_{\rm NLTE}$\\
\hline
C\,(CH)        & 8.39 &          &  0.05       & 0.11    &  -- &\\
N\,(CN)        & 7.78 &          &  <0.10      & --      &  -- &\\
\ion{O}{i}     & 8.66 &          &  1.10$^3$   & 0.15    & 2  & \\
\ion{Na}{i}    & 6.17 &     4.22 &  0.36       & 0.12    & 2   &$-$0.13\\
\ion{Mg}{i}    & 7.53 &     5.77 &  0.28       & 0.09    & 9   &0.28\\
\ion{Al}{i}    & 6.37 &     3.86 &  $-$0.20    & 0.11    & 2  & 0.06\\
\ion{Si}{i}    & 7.51 &     5.48 &  0.40       & 0.10    & 1   &0.34\\
\ion{K}{i}     & 5.08 &     5.08 &  0.24$^3$   & 0.10    & 1  & \\
\ion{Ca}{i}    & 6.31 &     4.53 &  0.39       & 0.12    & 12  &0.49 \\       
\ion{Sc}{ii}   & 3.05 &     0.87 &  $-$0.07    & 0.08    & 3  & \\         
\ion{Ti}{i}    & 4.90 &     3.10 &  $-$0.19    & 0.23    & 4   &\\
\ion{Ti}{ii}   & 4.90 &     3.01 &  $-$0.05    & 0.08    & 16  &$-$0.12\\
\ion{V}{i}     & 4.00 &     1.85 &  $-$0.02    & 0.08    & 8   &\\
\ion{Cr}{i}    & 5.64 &     3.23 &  $-$0.27    & 0.11    & 6   &\\
\ion{Mn}{i}    & 5.39 &     2.68 &  $-$0.25    & 0.08    & 3   &\\
\ion{Fe}{i}    & 7.45 &     5.31 &  $-$2.31$^4$& 0.06    &231  & $-$2.27\\
\ion{Fe}{ii}   & 7.45 &     5.16 &  $-$2.38$^5$    & 0.09    & 7   & \\
\ion{Co}{i}    & 4.92 &     2.83 &  0.08    & 0.12    & 6   & \\
\ion{Ni}{i}    & 6.23 &     3.84 &  $-$0.37    & 0.13    & 5    &\\
\ion{Cu}{i}    & 4.21 &     -- &  <$-$0.40& --    & --   &\\
\ion{Zn}{i}    & 4.60 &     3.84 &     0.17    & 0.13    & 2   & \\
\ion{Sr}{ii}   & 2.92 &     0.12 &     0.12    & 0.09    & 2   & \\
\ion{Y}{ii}    & 2.21 &     0.63 &  $-$0.63    & 0.09    & 4   & \\
\ion{Zr}{ii}   & 2.59 &     0.24 &     0.24    & 0.09    & 3   & \\
\ion{Ba}{ii}   & 2.17 &     0.13&      0.13    & 0.10    & 3   & \\
\ion{Eu}{ii}   & 0.52 &     0.55&      0.03    & 0.10    & 2   & \\
\hline
\multicolumn{7}{l}{$^{1}$Solar abundances adopted from \citet{asp05}} \\
\multicolumn{7}{l}{$^{2}$LTE and NLTE ratios are referred to [Fe/H]$_{\rm LTE}$ and [Fe/H]$_{\rm NLTE}$.} \\
\multicolumn{7}{l}{$^{3}$Abundance derived from the APOGEE spectrum and not used here.} \\
\multicolumn{7}{l}{$^{4}$[Fe/H] from \ion{Fe}{i} is given instead of [X/Fe]} \\
\multicolumn{7}{l}{$^{4}$[Fe/H] from \ion{Fe}{ii} is given instead of [X/Fe]} \\
\end{tabular}}
\end{table}